\documentclass[a4paper,fleqn,usenatbib]{mnras}

\usepackage[T1]{fontenc}
\usepackage{ae,aecompl}

\usepackage{graphicx}    % Including figure files
\usepackage{amsmath}    % Advanced maths commands
\usepackage{amssymb}    % Extra maths symbols

\title[Star formation history of the Galactic bulge]{Star formation history of the Galactic bulge from deep HST imaging of low reddening windows}

\author[E. J. Bernard et al.]{%
Edouard J. Bernard,$^{1}$\thanks{E-mail: ebernard@oca.eu (EJB)}
Mathias Schultheis,$^{1}$
Paola Di Matteo,$^{2}$
Vanessa Hill,$^{1}$
\newauthor
Misha Haywood,$^{2}$
Annalisa Calamida$^{3}$
\newauthor
\\
$^{1}$Universit\'e C\^ote d'Azur, OCA, CNRS, Lagrange, France\\
$^{2}$GEPI, Observatoire de Paris, CNRS, Universit\'e Paris Diderot, 5 place Jules Janssen, 92190, Meudon, France\\
$^{3}$National Optical Astronomy Observatory -- AURA, 950 N Cherry Avenue, Tucson, AZ, 85719, USA
}

\date{Accepted XXX. Received YYY; in original form ZZZ}

\pubyear{2018}

\begin{document}
\label{firstpage}
\pagerange{\pageref{firstpage}--\pageref{lastpage}}
\maketitle

\begin{abstract} % 250 words

Despite the huge amount of photometric and spectroscopic efforts targetting the Galactic bulge over the past few years, its age distribution remains
controversial owing to both the complexity of determining the age of individual stars and the difficult observing conditions.
Taking advantage of the recent release of very deep, proper-motion-cleaned colour--magnitude diagrams (CMDs) of four low reddening windows obtained with the \emph{Hubble Space Telescope (HST)}, we used the CMD-fitting technique to calculate the star formation history (SFH) of the bulge at $-2\degr > b > -4\degr$ along the minor axis.
We find that over 80~percent of the stars formed before 8~Gyr ago, but that a significant fraction of the super-solar metallicity stars are younger than this age. Considering only the stars that are within reach of the current generation of spectrographs (i.e.\ $V\la$~21), we find that 10~percent of the bulge stars are younger than 5~Gyr, while this fraction rises to 20--25~percent in the metal-rich peak.
The age--metallicity relation is well parametrized by a linear fit implying an enrichment rate of $dZ/dt \sim 0.005\ {\rm Gyr^{-1}}$. Our metallicity distribution function accurately reproduces that observed by several spectroscopic surveys of Baade's window, with the bulk of stars having metal-content in the range [Fe/H]$\sim-$0.7 to $\sim$0.6, along with a sparse tail to much lower metallicities.

\end{abstract}


%%%%%%%%%%%%%%%%% BODY OF PAPER %%%%%%%%%%%%%%%%%%

\section{Introduction}

The bulge is probably the most studied structural component of our Galaxy, yet due to its complexity and to difficult observing conditions -- stellar crowding, depth effect, interstellar reddening, foreground disc contamination -- its true nature is still a matter of debate.
It has long been thought of as very old and rather metal-rich \citep[e.g.][]{ren99}, a preconception motivated by both its apparent similarity to elliptical galaxies and by early stellar abundance patterns suggesting a short star formation and metal enrichment timescale \citep[e.g.][]{mcw94}.
And yet, it was recognised early on that the central concentration in galaxies are not always classical (i.e.\ merger built) bulges, but can result from the secular evolution of the disc, where disc instability and bar buckling lead to the formation of so-called pseudobulges \citep[see, e.g.][for a review]{kor04}.
Such pseudobulges retain the memory of their disc origins in that they tend to be rotationally dominated, be disky or boxy/peanut shaped, and of course share some of the stellar populations with the inner disc.

It is therefore reasonable to expect to find stars with a broad range of ages in the bulge. And yet, despite mounting evidence of the presence of younger stars \citep[e.g.][]{van03,cat16}, conclusively proving their actual youth and/or bulge membership has been challenging. Not only is determining the age of individual stars hard \citep{sod10}, but the few young A--F stars observed in the direction of the bulge cannot easily be distinguished from either foreground disc contaminants or blue straggler stars \citep[e.g.][]{zoc03,cla11}.
However, recent photometric data have brought solid evidence of the existence of young bulge stars. For example, a series of papers \citep[and references therein]{ben17} focusing on high-resolution spectroscopy of a sample of 90 microlensed dwarfs found that about one third of the super-solar metallicity stars are younger than 8 Gyr. \citet{val15} also analyzed independently this sample in order to address various sources of systematic uncertainties, and basically confirmed the findings of Bensby et al. Further, \citet{hay16a} presented a comparative analysis of a deep \emph{Hubble Space Telescope (HST)} colour--magnitude diagram (CMD) of a bulge field and found that, assuming any reasonable age--metallicity relation (AMR), the narrow width of the observed main-sequence turn-off (MSTO) and significant spectroscopic metallicity spread implies that there \emph{must} be a wide range of ages -- and that all the super-solar metallicity stars must be younger than $\sim$8~Gyr.

\begin{figure}
 \includegraphics[width=8.5cm]{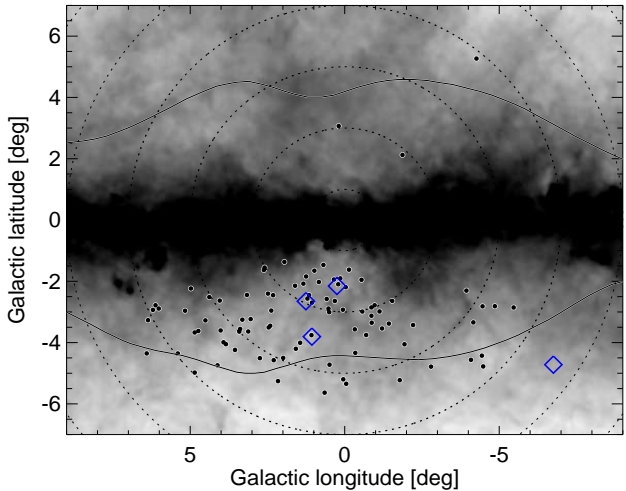}
 \caption{Location of the four HST fields (blue squares) in the bulge, which is represented by the solid lines, overlaid on the reddening map from \citet{sch98}. The microlensed dwarfs studied by \citet{ben17} are shown as filled circles.}
 \label{fig01}
\end{figure}

Here we extend the analysis of \cite{hay16a} by using the technique of synthetic CMD-fitting to provide a more detailed and robust quantitative analysis of this and other nearby \emph{HST} fields. We exploit the recently released deep CMDs of several low-reddening windows in the bulge to quantify its star formation history (SFH) using the CMD-fitting technique, which has been extensively validated in studies of nearby Local Group galaxies.
The paper is structured as follows: in Section~\ref{sec:2} we present the dataset upon which our analysis is based, as well as the steps taken to discriminate between bulge and foreground disc stars. The CMD-fitting method and the resulting SFHs are described in Sections~\ref{sec:3.1} and \ref{sec:3.2}, respectively. Our main results are presented in Section~\ref{sec:4}, where we compare our measured AMR, fraction of genuinely young stars, and present-day metallicity distribution function (MDF) with the results from the recent literature. A summary of the main conclusions closes the paper in Section~\ref{sec:5}.

\section{The dataset}~\label{sec:2}

This work is based on the public version 2 high-level science products from ``The WFC3 Galactic Bulge Treasury Program: Populations, Formation History, and Planets'' \citep[GO-11664;][]{bro09}\footnote{https://archive.stsci.edu/prepds/wfc3bulge/}, which covers four fields in low-reddening windows of the Galactic bulge labeled \emph{SWEEPS, Stanek, Baade}, and \emph{OGLE29}. The field locations are summarized in Table~\ref{tab:1} and shown in Figure~\ref{fig01} in relation with the bulge and a sample of microlensed dwarfs for which accurate ages are known \citep{ben17}.
For each field, they provide WFC3 photometry, astrometry, proper motions (PMs), and co-added (drizzled) images. The photometry is given in five bands: $F390W$, $F555W$, $F814W$, $F110W$, and $F160W$ in the STMAG photometric system, though in this work we only used UVIS' $F555W$ ($\sim V$) and $F814W$ ($\sim I$) filters.
As the isochrone set used in the SFH calculations (see section \ref{sec:3}) is provided in VEGAMAG, we converted the magnitudes from STMAG using the zero-points from \citet{deu17}\footnote{http://www.stsci.edu/hst/wfc3/documents/ISRs/WFC3-2017-14.pdf}.

\begin{figure}
 \includegraphics[width=8.5cm]{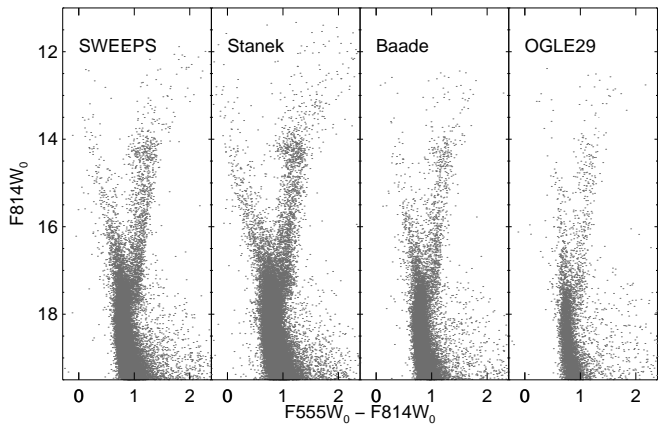}
 \caption{CMD of the 4 \emph{HST} fields in the bulge. Only stars brighter than $F814W_0=20$ were considered (our full sample), while the photometry is actually about 5 magnitudes deeper.}
 \label{fig02}
\end{figure}

For the reddening correction we started with the values provided in \citet{bro09}, increased by about 6~percent to adopt $E(B-V) = 0.55$ for Baade's window as in, e.g., \citet{sum04,kun08,zoc08}. We then refined these estimates by finding the values that minimized the difference in both luminosity and colour functions between the fields; this only resulted in changes of at most one percent. The final, adopted colour excesses are listed in Table~\ref{tab:1}. Given the small field-of-view of the \emph{WFC3} ($162\arcsec\times162\arcsec$), we neglect any differential extinction variation inside each field. The CMDs, dereddenned using the extinction coefficients from \citet{sch11} and assuming $R_V$=3.1, are shown in Figure~\ref{fig02}. We note that for this work we have only used stars brighter than $F814W_0=20$, while the photometry is actually about five magnitudes deeper; we are therefore only using stars with high signal-to-noise ratios. For the SFH calculations, we also transformed the photometry to absolute magnitudes assuming a bulge distance of 8~kpc \citep[e.g.][and references therein]{val17}.

\begin{table}
\centering
\caption{WFC3 Galactic Bulge Treasury fields}
\label{tab:1}
\begin{tabular}{lccc}
\hline
                & $l$      & $b$     & $E(B-V)$ \\
Field name      & (deg)    & (deg)   &          \\
\hline
SWEEPS          & +1.2559  & -2.6507 & 0.66     \\
Stanek          & +0.2508  & -2.1531 & 0.90     \\
Baade           & +1.0624  & -3.8059 & 0.55     \\
OGLE29          & -6.7532  & -4.7195 & 0.58     \\
\hline
\end{tabular}
\end{table}

\begin{figure}
 \includegraphics[width=8.cm]{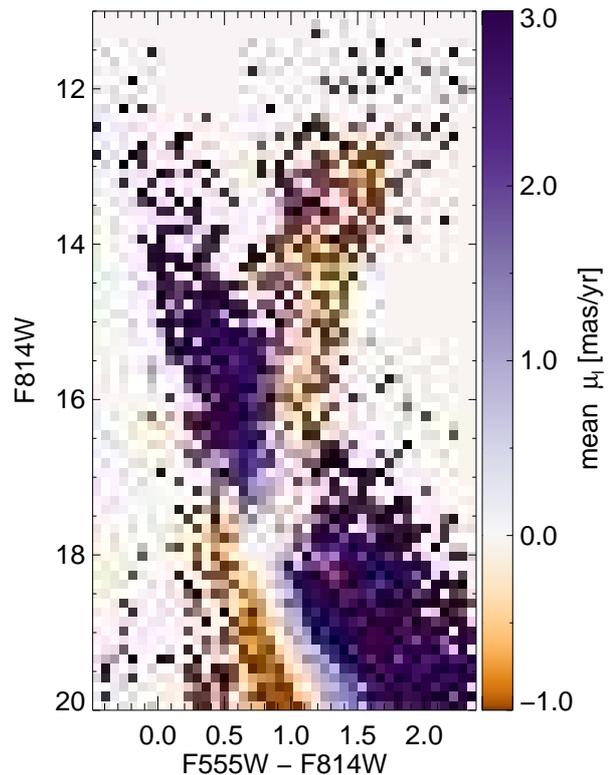}
 \caption{Hess diagram of the combined \emph{SWEEPS+Stanek+Baade+OGLE29} CMD, colour-coded by mean proper motion $\mu_l$. The contamination from the foreground disc stars is clearly visible in dark blue.}
 \label{fig03}
\end{figure}

\begin{figure}
 \includegraphics[width=8.5cm]{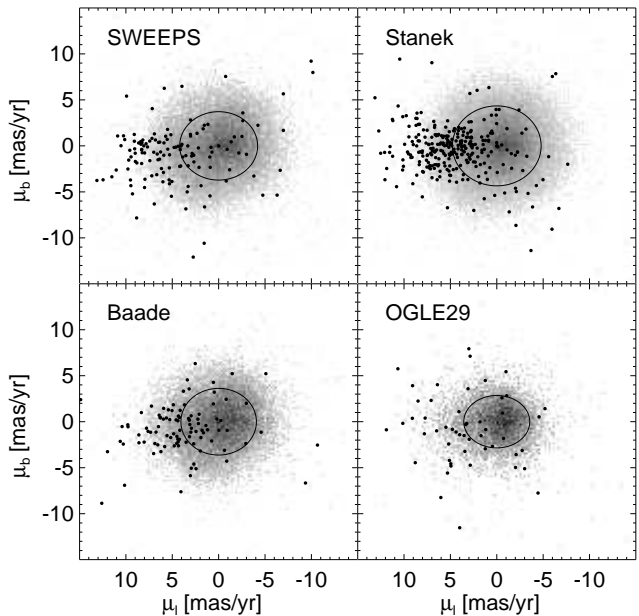}
 \caption{Proper motion distributions in each bulge field. Larger dots correspond to main-sequence stars brighter than $F814W_0\sim16$, and the ellipses represent the standard deviation in $\mu_l$ and $\mu_b$.}
 \label{fig04}
\end{figure}

An important aspect of this dataset is the fact that it includes precise PMs for all the stars, which we can use to separate bulge stars from foreground disc stars. They are based on repeated observations with \emph{HST-WFC3}. The time baseline for \emph{SWEEPS} is 2,266 days, while it is about 750 days for the other three fields\footnote{https://archive.stsci.edu/pub/hlsp/wfc3bulge/README}. The PMs are provided in units of pixels in equatorial coordinates, and were transformed to mas yr$^{-1}$ in Galactic coordinates following e.g.\ \citet{pol13}.
To illustrate the potential for bulge/disc separation, Figure~\ref{fig03} shows the mean proper motion along the Galactic longitude in colour and magnitude bins for the combined \emph{SWEEPS+Stanek+Baade+OGLE29} CMD, where the rotating disc component is clearly visible in dark blue.
The PMs of the full samples are shown as gray scale in Figure~\ref{fig04}. To highlight the contribution of disc stars, we overplotted the location of bright main sequence stars ($F814W_0\la16$ and $(F555W-F814W)_0<0.8$) as larger symbols. The disc component is clearly visible at $\mu_l > 0$ and $\mu_b \sim 0$. While the catalogues released by the \emph{Treasury Program} do not include uncertainties on the proper motions, \citet{cla08} have used a similar \emph{HST} dataset with a 2-year baseline to show that proper motions for stars with ${\rm F814W}<20$ can be measured to better than 0.3 ${\rm mas\ yr}^{-1}$. The dispersion in proper motion shown in Figure~\ref{fig04} is therefore fully dominated by the intrinsic motion of these relatively bright stars.

Following \citet{cla08}, we discriminate between ``disc'' and bulge stars based on their longitudinal PM. The decomposition is shown in Figure~\ref{fig05}, where the fraction of stars with disc-like kinematics (blue curve) is labelled. It ranges from about 10~percent for the two fields nearest the Galactic center to 49~percent in \emph{OGLE29}, though we note that the bright MSTO stars are much more dispersed in the PM distribution of the latter field than in the other fields. It is also too sparsely populated once PM-cleaned of the disc contamination. For these reasons, in the remainder of this paper we only analyze the \emph{SWEEPS, Stanek}, and \emph{Baade} fields.

According to the decomposition shown in Figure~\ref{fig05}, a proper motion cut at $\mu_l < 3\ {\rm mas\ yr}^{-1}$ should reject most of the ``disc'' stars in these three fields, and contamination should be negligible with a more stringent cut at $\mu_l < -3\ {\rm mas\ yr}^{-1}$: the remaining fraction of ``disc'' stars after each cut indicated in each panel shows that this is indeed the case. We refer to these subsets as our \emph{Clean} and \emph{Cleanest} samples.
We have also checked that these cuts have a negligible impact on the SFH by applying the same cuts to the bulge model of \citet[see Appendix~\ref{A1}]{fra17}. The corresponding CMDs for the \emph{Baade} field are shown in Figure~\ref{fig06}. It reveals that the \emph{Clean} bulge CMD still contains a relatively prominent plume of stars brighter and bluer than the MSTO corresponding to stars younger than 6--8 Gyr old, and stars redder than the main-sequence at $F814W_0\ga17$ which are likely foreground (disc) dwarfs. On the other hand, these populations are almost entirely missing in the \emph{Cleanest} bulge CMD.

Finally, since the proper motion cleaned CMDs (and the resulting SFHs, see Appendix~\ref{C1}) of the three fields are very similar to each other, we combined them to obtain significantly more populated CMDs, and therefore more robust SFHs. The CMDs are shown in Figure~\ref{fig07}.

\section{Star formation history calculations}\label{sec:3}

\subsection{Methodology}\label{sec:3.1}

The SFH calculations have been carried out using the technique of synthetic CMD-fitting following the methodology presented in, e.g.\ \citet{mon10,ber12,ber15a}.
It involves fitting the
observed data with synthetic CMDs to extract the linear combination of
simple stellar populations (SSPs) -- i.e.\ each with small ranges of
age ($\leq$1~Gyr) and metallicity ($\sim$0.2~dex) -- which provide the
best fit; the amplitudes of which correspond to the rates of star formation as
a function of age and metallicity. The minimization is done with the algorithm described in \citet{ber15b}.

A new feature of the algorithm is the possibility to use CMDs (empirical or synthetic) corresponding to foreground/background populations as additional SSPs.
This significantly improves the fit when the contamination is important, since the algorithm does not attempt to fit the contaminants using SSPs at the distance of the population of interest, and therefore leads to a cleaner AMR. Here, we have selected the observed stars with clear disc-like rotation ($\mu_l > 6\ {\rm mas\ yr}^{-1}$) as the contaminant population. The corresponding CMD, shown in the right panel of Figure~\ref{fig08}, harbors a prominent main-sequence with significant magnitude spread as a consequence of their large range of heliocentric distances.
Since this foreground population is actually a subset of the `full' CMD, we have only used it in the fitting of the `clean' and `cleanest' CMDs.

The synthetic CMD from which we extracted the SSP CMDs, shown in the middle panel of Figure~\ref{fig08}, was generated with the BaSTI stellar evolution library \citep{pie04}. We adopted a \citet{kro93} initial mass function, and assumed a binary fraction of 40~percent and a mass ratio $q$ uniformly distributed between 0.1 and 1 \citep[e.g.][]{tok14}. It contains 30 million stars down to a mass of 0.7~M$_{\sun}$ and was generated with a constant SFR over wide ranges of age and metallicity: 0 to 14 Gyr old and
0.0004 $\leq$ Z $\leq$ 0.05 (i.e.\ -1.7 $\leq$ [Fe/H] $\leq$ 0.5, assuming
Z$_{\sun}$ = 0.0198 and no $\alpha$-enhancement; \citealt{gre93}).
We also carried-out the CMD-fitting with the $\alpha$-enhanced models ($\rm{[Fe/H]}\sim0.4$) of the BaSTI library \citep{pie06}, which tend to yield ages that are $\sim$10--15~percent younger. While our conclusions are unchanged, the fractions of stars younger than 8~Gyr in the bulge that we calculate in Section~\ref{sec:4.2} may actually be slightly underestimated if the age of the older, $\alpha$-enhanced stars is overestimated.

Before the observations can be compared to the model, it is necessary to simulate realistic incompleteness and photometric errors due to the observational effects on the synthetic CMD. These are typically estimated using artificial stars tests on the original images \citep[e.g.][]{gal99}. However, these tests were not provided with the WFC3 catalogues. On the other hand, incompleteness is only significant in very crowded fields or near the detection limit; since there are only between $\sim$153,000 and $\sim$213,000 stars in each field and we are using stars $\sim$5 magnitudes brighter than the detection limit, we can safely assume that there are no variation of completeness with magnitude or colour. Photometric errors in the magnitude range considered are negligible ($<$0.01) for the \emph{SWEEPS} and \emph{Baade} fields, but $\sigma_{555}\sim$\,0.042 and $\sigma_{814}\sim$\,0.016 for the \emph{Stanek} field. We therefore applied random, Gaussian magnitude offsets with the stated standard deviations to the synthetic stars for the SFH calculations of the latter field.

\begin{figure}
 \includegraphics[width=8.5cm]{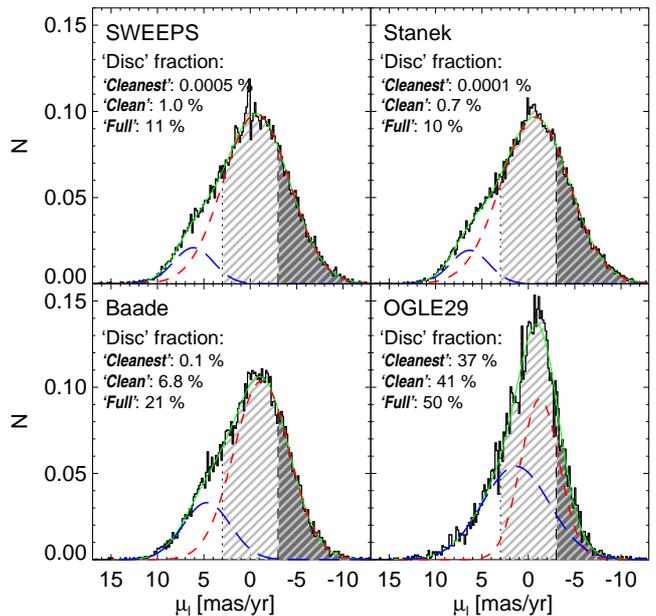}
 \caption{Longitudinal proper motion distributions of the full sample in each bulge field. Red and blue lines show the Gaussian decomposition in `bulge' and `disc' components, respectively, and the sum of the two is shown in green. The PM cuts (dotted and dashed vertical lines) define our \emph{clean} and \emph{cleanest} samples shown as light and dark gray shading, respectively. The fraction of `disc' stars in each sample, from the Gaussian fits, is indicated in each panel.}
 \label{fig05}
\end{figure}

In addition, for all the fields we added further offsets in magnitude to take into account the bulge thickness along the line of sight.
From the Besan\c con \citep{rob03} and Trilegal \citep{gir05} models, we find that in the \emph{SWEEPS} field the bulge is 1.19 and 0.95~kpc thick, respectively, with less than 5~percent difference between the three fields according to the former model. The negligible variations between these fields despite their different latitudes is a consequence of the boxy shape of the bulge in this region due to the presence of the bar. We therefore used a Gaussian dispersion with standard deviation of 1~kpc, corresponding to a spread of $\sim$0.27~mag for a bulge distance of 8~kpc. This value is commonly used for the analysis of bulge data \citep[e.g.][]{val13,hay16a}. Using a lower thickness (e.g.\ $\sigma=$~0.75~kpc) led to worse SFH fits.

\begin{figure}
 \includegraphics[width=8.5cm]{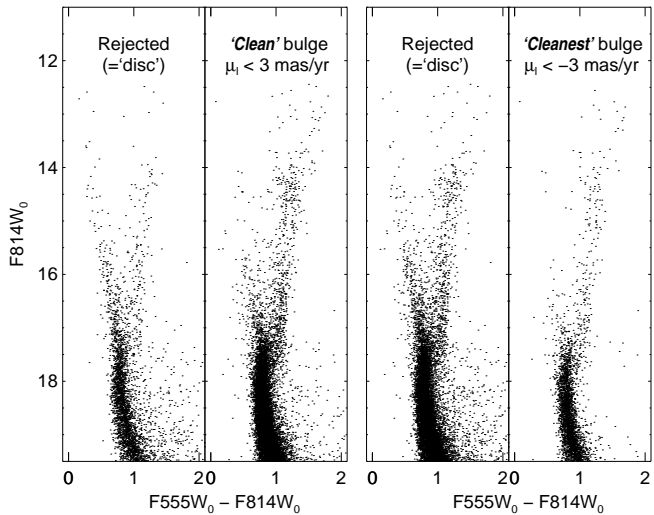}
 \caption{Proper-motion cleaning of the \emph{Baade} field CMD. The rejected and retained stars for the \emph{Clean} (left) and \emph{Cleanest} (right) samples are shown.}
 \label{fig06}
\end{figure}

The comparison between the observed and synthetic CMDs is performed using the number of stars in small colour--magnitude boxes. No a priori constraint on the AMR is adopted: the algorithm solves for both ages
and metallicities simultaneously within the age-metallicity space covered
by our SSPs. The goodness of the coefficients in the linear combination
is measured through a Poissonian equivalent to the $\chi^2$ statistic \citep[$\chi_P^2$, see e.g.][]{dol02}, which is minimized using the Python-SciPy implementation of the Truncated Newton Conjugate-Gradient algorithm \citep{nas84}. These coefficients are directly proportional to the star formation rate of their corresponding SSPs.

To sample the vast parameter space, the $\chi_P^2$ minimization is
repeated numerous times for each field after shifting the bin
sampling in both age--metallicity and colour--magnitude space. In
addition, the observed CMD is also shifted with respect to the synthetic
CMDs in order to account for uncertainties in photometric zero-points,
distance, and mean reddening. Finally, the uncertainties on the SFRs
were estimated as described in \citet{hid11}. The
total uncertainties are assumed to be a combination in quadrature of
the uncertainties due to the effect of binning in the colour--magnitude
and age--metallicity planes, and those from the statistical sampling in the observed CMD.

\subsection{Star formation histories}\label{sec:3.2}

The resulting SFHs for the combined \emph{SWEEPS+Stanek+Baade} CMDs are shown in Figure~\ref{fig09}. For completeness, the SFHs of each individual field are presented in Appendix~\ref{C1}. For each PM-cleaning level, the top and bottom panels represent the cumulative fraction of the mass of stars formed and the metallicity, respectively, as a function of time.

\begin{figure}
 \includegraphics[width=8.2cm]{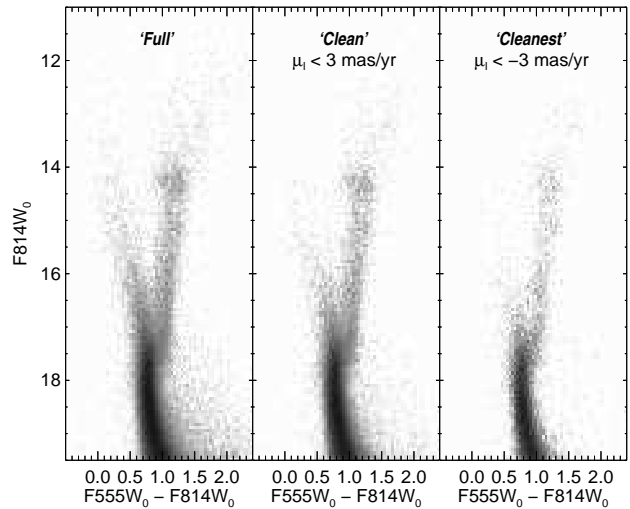}
 \caption{Three levels of proper-motion cleaning for the combined \emph{SWEEPS+Stanek+Baade} CMD.}
 \label{fig07}
\end{figure}

Overall, in all cases the SFH shows a globally old population, with over 50~percent stars formed before 10~Gyr ago, and 80~percent of stars formed before $\sim$8~Gyr ago (i.e.\ a redshift of $\sim$1). However, there also appears to be some residual star formation until about 1~Gyr ago.
The \emph{Clean} (middle) and \emph{Cleanest} (right) samples present an older mean age than the \emph{Full} (left) sample, as expected from the vanishing contamination by foreground disc stars thanks to both PM-cleaning and the fitting of the remaining contaminants. In fact, their SFH are very similar to each other, which shows that simultaneously fitting the foreground population allows us to recover the intrinsic SFH of the bulge as if the disc contamination was not present.

\begin{figure*}
 \includegraphics[width=12cm]{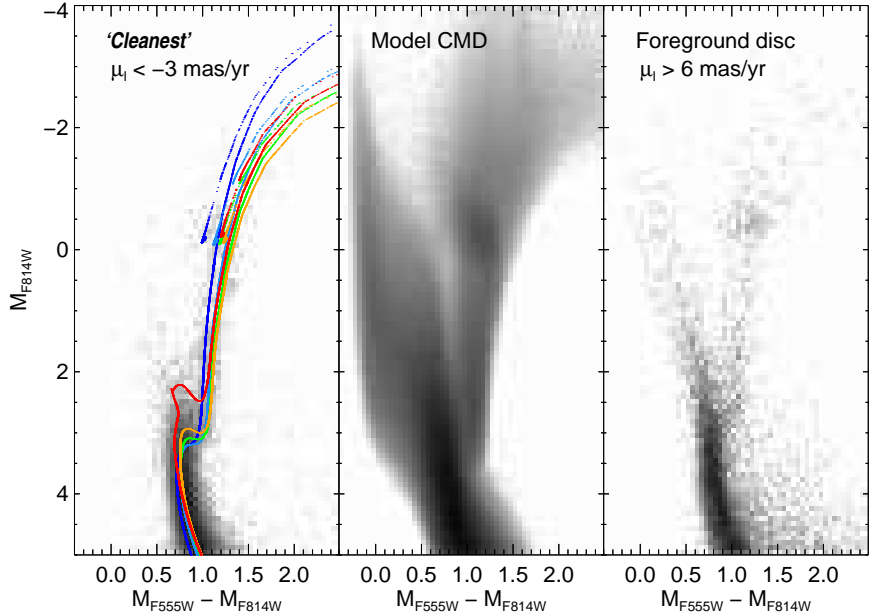}
 \caption{\emph{Left:} Hess diagram of the combined \emph{SWEEPS+Stanek+Baade} CMD where isochrones corresponding to the AMR determined in section~\ref{sec:4.1} have been overplotted: for ages of 13 (dark blue), 11 (light blue), 9 (green), 7 (orange), and 3.5 (red) Gyr we have used metallicities of Z = 0.0087, 0.0194, 0.0300, 0.0400, and 0.0360 (i.e.\ [Fe/H]$\sim$-0.31, 0.05, 0.25, 0.40, and 0.34).
 \emph{Middle:} Model CMD from which the SSPs have been extracted (see section~\ref{sec:3.1}).
 \emph{Right:} Hess diagram of the CMD corresponding to the foreground population, selected using $\mu_l > 6\ {\rm mas\ yr}^{-1}$.}
 \label{fig08}
\end{figure*}

This is consistent with the presence of a plume of blue stars above the MSTO (see, e.g., the right panel in Figure~\ref{fig07}). However, purely old stellar systems are known to harbor so-called blue straggler stars (BSS) in this region \citep[e.g.][]{mom07,mon12,fer12}, i.e.\ old stars that have been rejuvenated through mass-transfer in a binary system or through stellar collisions. To make sure that the recent star formation episode we measure is not affected by these stars, we repeated the SFH calculations without including stars brighter than $F814W_0 = 17.2$ and bluer than $(F555W - F814W)_0 = 0.85$. The resulting SFH is indistinguishible from that calculated with the blue plume, suggesting that actually young stars are responsible for most of the SFR since about 7~Gyr ago.

Our SFH is a good match to the SFH of the inner disc obtained by \citet{sna15} from fitting a chemical evolution model to a large sample of stellar abundances. It also presents a strong initial SFR for the first 4--5 Gyr before maintaining low-level star formation until the recent past. While their median age is about 2~Gyr younger than ours, we emphasize that both methods are based on models with different assumptions, such that difference in the \emph{absolute} ages are expected \citep[see, e.g.][for a comparison of the SFHs obtained with six different stellar evolution libraries]{ski17}.
In addition, our calculations with $\alpha$-enhanced models produced slightly younger ages that are in better agreement with the results of \citet{sna15} (see Section~\ref{sec:3.1}).

In the bottom panels showing the AMRs, the SSPs contributing most to the CMDs appear as darker shades of gray, while the blue points and bars represent the mean metallicity and dispersion in each age bin (averaged in Z rather than in [Fe/H]). In all cases the AMR indicates a quick metal enrichment, reaching super-solar metallicity about 8 Gyr ago but with little evolution since then. Note that in the left panel the metallicity of the stars younger than 8 Gyr seems to be strongly limited by the edge of the grid. This is actually an artefact of the method due to the assumption that all the stars are located at the same distance. Since the disc contamination lies mostly in the foreground, the young stars appear brighter and/or redder (see e.g.\ Figure~\ref{fig03}), which the algorithm can only fit with stars of very high metallicity. This effect is significantly reduced in the other two panels since the foreground has been fitted correctly.

In the following section we use these SFHs to quantify in more details the AMR of the bulge, estimate the fraction of genuinely young stars, and determine the present-day MDFs; we also compare these with values from the literature.

\begin{figure*}
 \includegraphics[width=5.8cm]{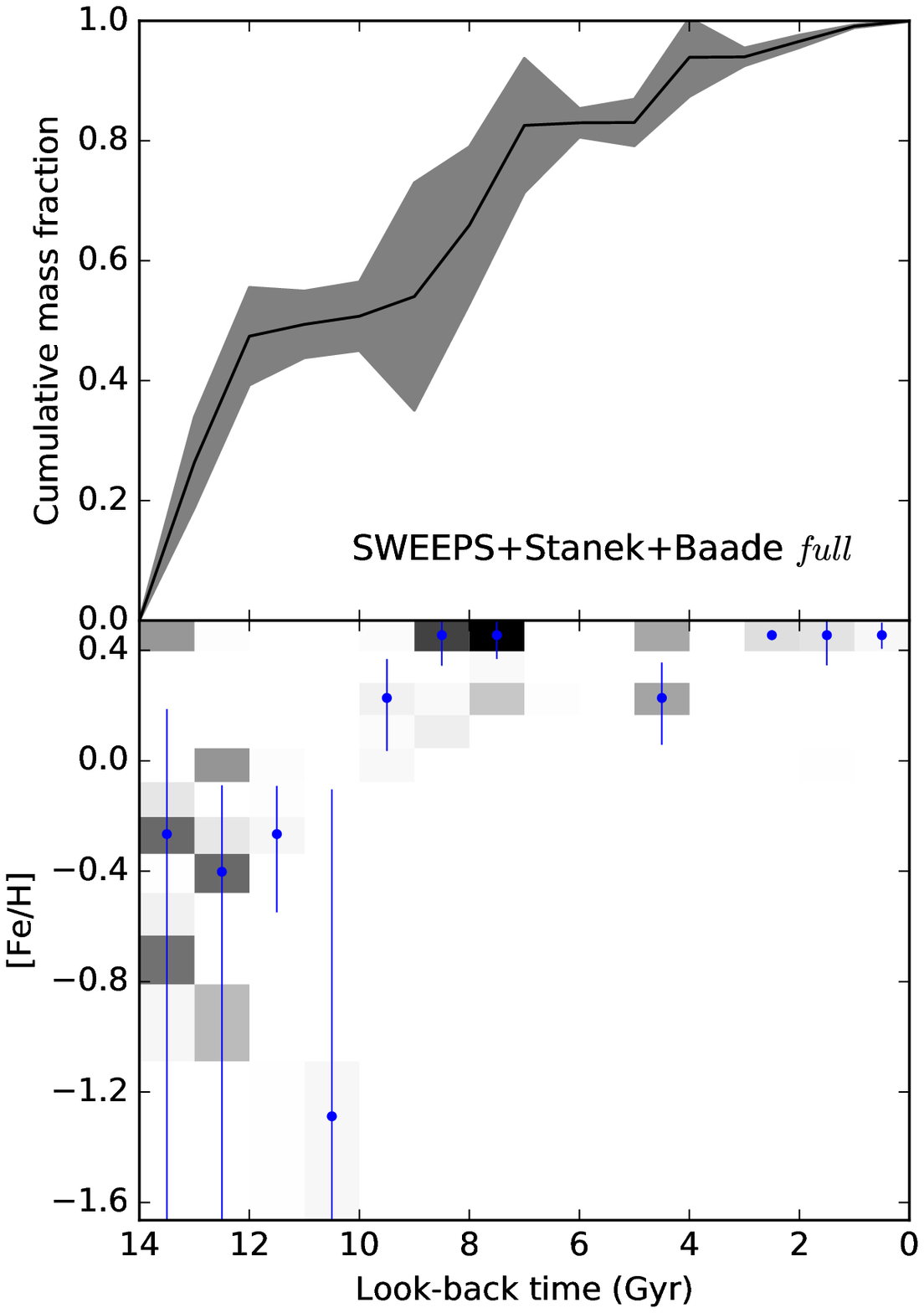}
 \includegraphics[width=5.8cm]{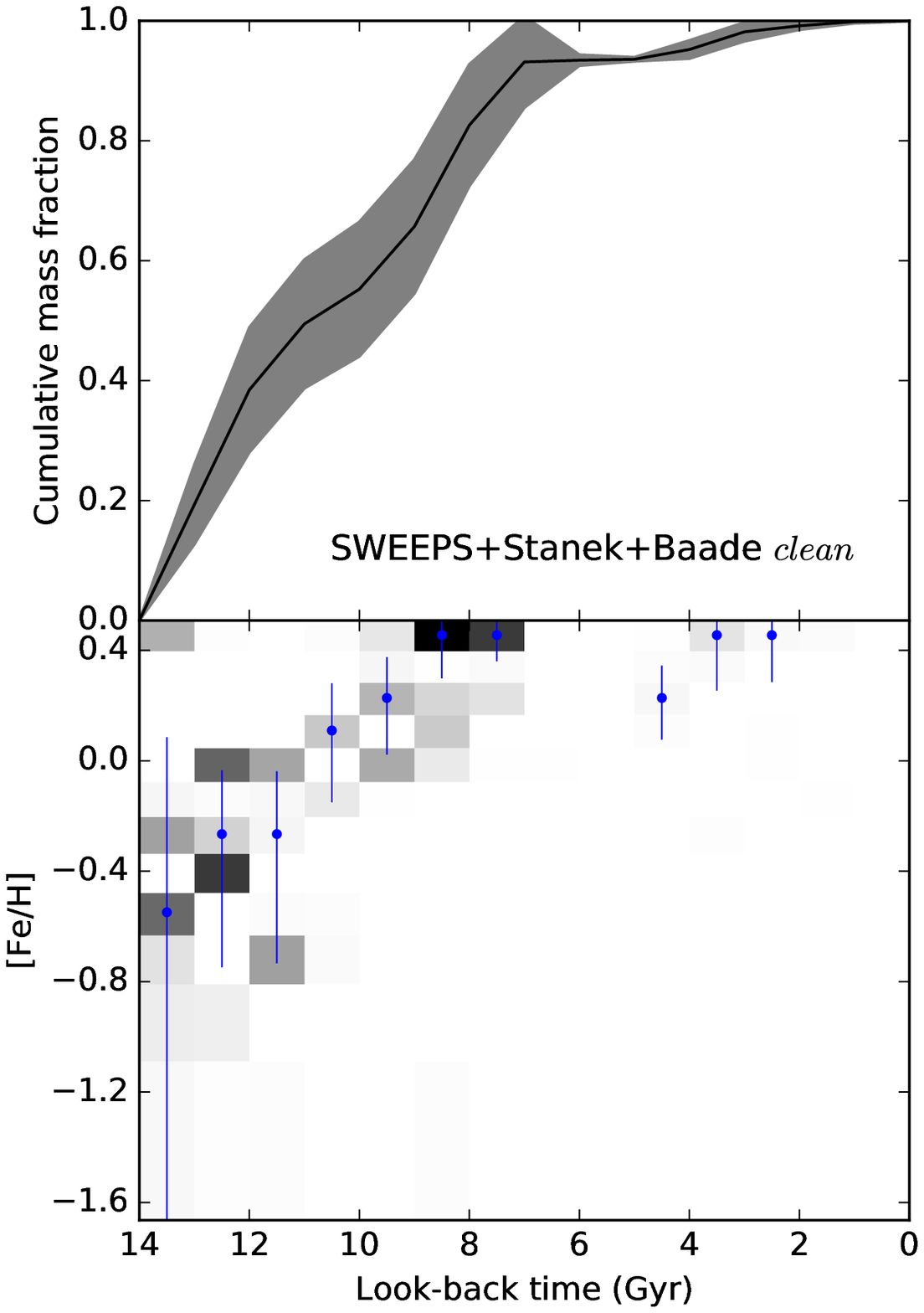}
 \includegraphics[width=5.8cm]{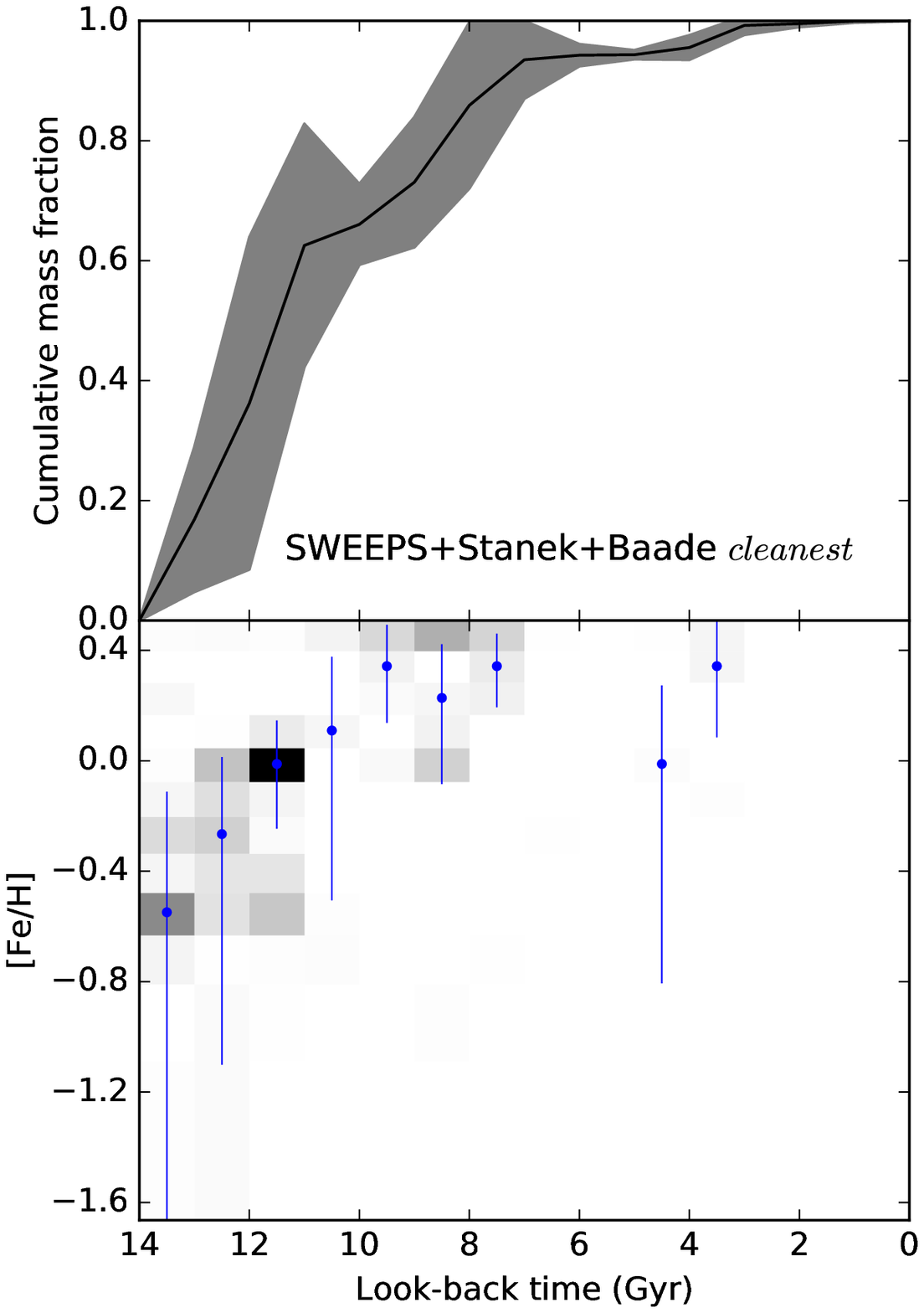}
 \caption{SFHs of the combined \emph{SWEEPS+Stanek+Baade} fields from the full (left), clean (middle), and cleanest (right) CMDs. The top and bottom panels show the evolution of the total mass of stars formed and of the metallicity, respectively.
 In the bottom panels the SSPs contributing most to the CMDs appear as darker shades of gray; the blue points and bars represent the mean metallicity and dispersion in each age bin.}
 \label{fig09}
\end{figure*}

%%%%%%%%%%%%%%%%%%%%%%%%%%%%%%%%%%%%%%%%%%%%%%%%%%
\begin{table*}
\centering
\caption{Percentage of stars younger than the stated age in three metallicity bins for each proper-motion cleaning levels.}
\label{tab:2}
\begin{tabular}{lccccc}
\hline
 Younger than... & Cleaning level & [Fe/H]$\le$$-$0.5 & $-$0.5$<$[Fe/H]$\le$0 & 0$<$[Fe/H] & All [Fe/H] \\
\hline
          & CLEANEST &     6.3 $\pm$ 2.6 & 9.5 $\pm$ 1.7 & 32.9 $\pm$ 1.8 & 19.2 $\pm$ 1.1 \\
  8 Gyr   & CLEAN    &     2.8 $\pm$ 1.5 & 7.5 $\pm$ 1.2 & 37.5 $\pm$ 0.9 & 22.9 $\pm$ 0.6 \\
          & FULL     &     3.2 $\pm$ 1.3 & 1.3 $\pm$ 1.5 & 62.0 $\pm$ 0.7 & 40.9 $\pm$ 0.6 \\
\hline
          & CLEANEST &     4.0 $\pm$ 2.7 & 7.5 $\pm$ 1.7 & 13.5 $\pm$ 1.7 &  9.5 $\pm$ 1.1 \\
  5 Gyr   & CLEAN    &     1.9 $\pm$ 1.6 & 6.3 $\pm$ 1.3 & 15.6 $\pm$ 0.9 & 10.6 $\pm$ 0.6 \\
          & FULL     &     1.5 $\pm$ 1.3 & 1.3 $\pm$ 1.5 & 35.9 $\pm$ 0.7 & 23.8 $\pm$ 0.5 \\
\hline
\end{tabular}
\end{table*}

%%%%%%%%%%%%%%%%%%%%%%%%%%%%%%%%%%%%%%%%%%%%%%%%%%

\section{Results}\label{sec:4}

\subsection{The age-metallicity relation (AMR)}\label{sec:4.1}

\begin{figure}
 \begin{center}
 \includegraphics[width=8.5cm]{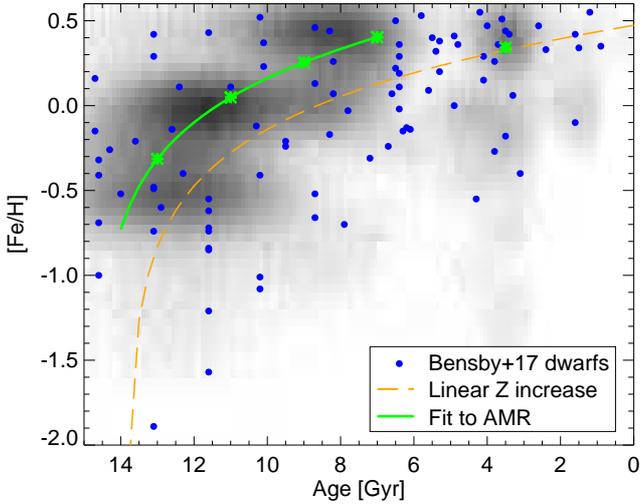}
 \end{center}
 \caption{Age--metallicity relation for MSTO stars currently alive, from the best-fit CMD of the \emph{cleanest} field, shown as gray-scale. The microlensed dwarfs from \citet{ben17} are shown as blue points, while the solid green and dashed orange lines are fits to the AMR of the \emph{cleanest} field and to the model AMR from \citet{hay16a}, respectively. The green asterisks along the mean AMR represent the five isochrones overplotted on Figure~\ref{fig08}.}
 \label{fig10}
\end{figure}

The AMRs shown in Figure~\ref{fig09} comprise all the stars that ever formed, while those determined from the age of individual stars \citep[e.g.][]{ben17} only include stars that are still alive today. To ease the comparison with observations we use the synthetic CMD corresponding to the SFH of the \emph{cleanest} sample.
While the best-fit CMD, by design, only contains approximately the same number of stars as in the observed CMD, we generated a much larger CMD ($\sim7.4\times10^5$ stars) with the same SFH to better sample the age--metallicity space.
We reproduced the selection function from \citet{ben17} by keeping only MSTO stars (i.e.\ with $16.5<F814W_0<19$), and applied small random offsets (10~percent in age, 0.1~dex in metallicity) to each star. The result is shown in Figure~\ref{fig10}, where we also overplot the microlensed dwarfs from \citet{ben17} as blue dots.
The dashed orange line is a linear fit to the model AMR used by \citet{hay16a}, and represents a metal enrichment from $Z=0$ to $Z=0.047$ (i.e.\ [Fe/H]$\sim$0.05) at a constant rate of $dZ/dt = 0.0034\ {\rm Gyr^{-1}}$. The solid green line is a similar fit to the AMR of the \emph{cleanest} CMD between 14 and 7~Gyr ago (blue points in Figure~\ref{fig09}). It implies a metal enrichment at a slightly higher rate of $dZ/dt \sim 0.005/ {\rm Gyr^{-1}}$, or $0.006\ {\rm Gyr^{-1}}$ if imposing $Z=0$ at $t=0$.

We find that the agreement between our AMR, shown as the gray-scale density plot, and the distribution of microlensed dwarfs is excellent. Both show a broad range of metallicities before 10~Gyr ago, and a more limited range of [Fe/H$\ga-$0.5 since then. Note, however, that the dwarfs seem to indicate a higher fraction of stars younger than $\sim$7~Gyr than our AMR, which is dominated by old populations; this is a consequence of the microlensing selection bias that favors the observation of young, metal-rich stars \citep[see][and section~\ref{sec:4.2}]{ben13}.

In addition, according to both our AMR and the microlensed dwarfs the mean metallicity has not evolved much since about 8~Gyr ago, most likely a result of the very low SFR, possibly combined with the infall of metal-poor gas in this part of the bulge. Interestingly, for [Fe/H]$>$0 there is a hint of age-bimodality, with a peak at 7--10 Gyr which is due to the end of the main formation episode, and a more recent formation episode from $\sim$2--5 Gyr. A similar bimodality at high metallicity has been noted before by \citet{sch17} from their analysis of APOGEE \citep{maj17} giants in Baade's window.

\subsection{On the fraction of young stars in the bulge}\label{sec:4.2}

The bulge has traditionally been thought of as a purely old component formed during an early monolithic collapse \citep[see, e.g.][for a historical perspective]{wys97}. However, recent observations have shown strong evidence of the presence of a significant number of young stars within the bulge. Here, we use the SFHs calculated from the deep CMDs to estimate the fraction of young stars ($\la$5--8 Gyr) in the region of the bulge covered by the \emph{HST} fields.

\begin{figure}
 \includegraphics[width=8.5cm]{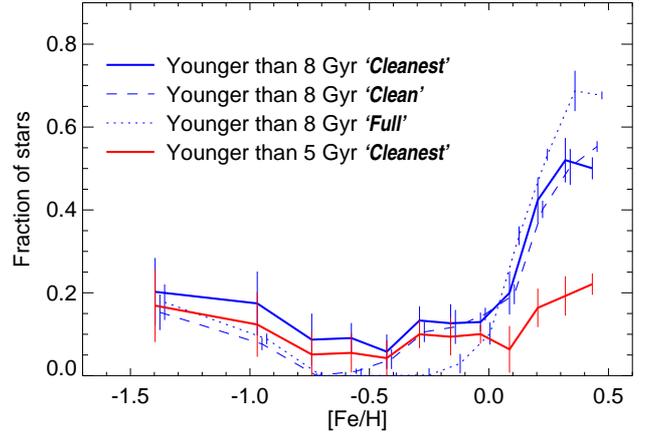}
 \caption{Fraction of stars younger than 8 Gyr and 5 Gyr for the 3 decontamination levels. Only stars near the MSTO (i.e.\ with 2 $<$ F814W $<$ 4.5) have been used for comparison with the sample of \citet{ben17}. The increase toward low metallicities is not significant given the very small number of stars in this regime.}
 \label{fig11}
\end{figure}

For this, we used the large synthetic CMD discussed above, in which we again selected the stars in the region of the MSTO -- i.e.\ with $16.5<F814W_0<19$ -- for comparison with the work of \citet{ben17}. The advantage of using a CMD rather than the AMR itself is that it properly takes into account the limited lifetime of stars of given ages and metallicities. We then measured the fraction of stars younger than a given age as a function of metallicity, with 0.2~dex bins and a step of 0.1 dex. In Figure~\ref{fig11} we show the results for 8~Gyr as blue lines with different linestyles for the three PM-cleaning levels.
The lower fraction of young stars in the \emph{clean} and \emph{cleanest} samples is clearly visible at high metallicity, as expected from the vanishing fraction of foreground disc stars after the PM-cleaning and CMD-fitting.
This is also the case for the fraction of stars younger than 5~Gyr (see Table~\ref{tab:2}), so for clarity we only show the curve corresponding to the \emph{cleanest} CMD in red.
The increasing fraction toward low metallicities is not significant given the very small number of stars in this regime. In all cases, we find that the fraction of young stars is minor up to solar abundance, but becomes significant for higher metallicities.

The values shown in Figure~\ref{fig11}, averaged over broader metallicity bins, are summarized in Table~\ref{tab:2}. Our fractions of stars younger than 8~Gyr (6, 10, 33~percent) are in excellent agreement with the (corrected) values given in \citet[9, 20, and 35~percent;][]{ben17}, and confirm the conclusion of \citet{ben13} that the values uncorrected for their biased selection function are overestimated by about 50~percent.
However, given the shape of the cumulative mass fraction (CMF) shown in the right panel of Figure~\ref{fig09}, we caution against using 8~Gyr as the limit to define the fraction of young stars. Because of the steepness of the CMF at this age, small systematic age differences due to e.g.\ different stellar evolution libraries could shift the estimated fraction by over 10~percent. Instead we recommend using 5~Gyr, where the CMF is almost flat, as the limit.
While \citet{ben17} do not explicitly quote their fractions of young stars in each metallicity bins when using a 5~Gyr cut-off, rough estimates from their Figure~14 -- corrected for the $\sim$50~percent over-estimation -- indicate excellent agreement with our values given in Table~\ref{tab:2}.

Finally, we note that the colour--magnitude selection function has a small but systematic effect on the estimated fractions of young stars. For example, compared to using only stars at the MSTO, the fraction of stars younger than 8 Gyr is on average $\sim1$~percent larger if using RGB stars (i.e.\ with $(F555W - F814W)_0>0.8$ and $F814W_0<14.5$).

\subsection{Metallicity distribution function (MDF)}

\begin{figure}
 \begin{center}
 \includegraphics[width=8.5cm]{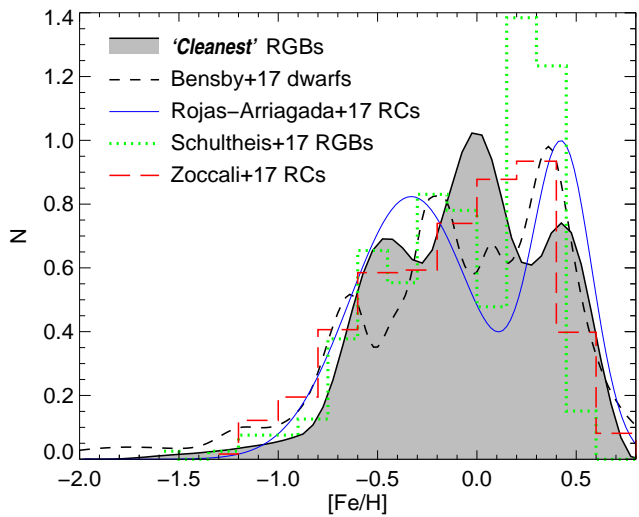}
 \end{center}
 \caption{Normalized metallicity distribution function from the RGB stars of the \emph{cleanest} CMD, compared to recent literature MDFs of nearby fields (see text for details).}
 \label{fig12}
\end{figure}

Figure~\ref{fig12} shows the present-day MDF obtained from the large, synthetic CMD corresponding to the SFH of the \emph{cleanest} CMD, after convolution with a kernel of $\sigma=0.15$\,dex. As a consequence of the smoothing, a small fraction of stars appear beyond the limit of our metallicity grid at [Fe/H]$>$0.5.
This suggests that even more metal-rich isochrones could have been used, although they are not presently available in the BaSTI stellar evolution library.
For comparison we also overlaid the MDF from recent spectroscopic surveys of stars in the vicinity of our fields. The dashed line represent the sample of microlensed dwarfs from \citet{ben17} already described above; the blue line is the double-Gaussian fit to the observed MDF from the 359 red clump (RC) giants in field $p1m4$ -- i.e.\ at ($l$,$b$) = ($1.00\degr$, $-3.97\degr$) -- of the Gaia-ESO survey \citep{roj17};
the dotted histogram is the MDF of the APOGEE sample of bulge RGB stars in Baade's window \citep{sch17}; finally, the MDF from the RC stars of fields $LRp0m2$ and $Baade's\ Window$ from the GIBS survey \citep{zoc17} is represented by the long dashed, red histogram.

While the detailed structure -- i.e.\ the location of the main peaks -- is different between the various MDFs, we note that the relatively small number of stars in each of these samples ($\la$600) may also lead to stochastic variations. The differences may in fact arise from using different stellar types as tracers (dwarfs, RC giants, cool RGBs), or be due to different metallicity calibrations. However, the large-scale distribution is fully consistent: most find that the bulk of stars have metal-content in the range [Fe/H]$\sim-$0.7 to $\sim$0.6, along with a sparse tail to much lower metallicities.

\section{Summary and Conclusions}\label{sec:5}

We have exploited the recently released deep \emph{HST} CMDs of several low-reddening windows along the minor axis of the Galactic bulge at $b\sim-3\degr$ to quantify its SFH using the CMD-fitting technique. We used the precise PMs afforded by the multi-epoch \emph{WFC3} observations to remove the foreground disc contamination, and made sure that this did not introduce any bias in the SFH by applying similar cuts to the bulge model of \citet{fra17}.
The quantitative SFH reveals a globally old bulge, with over 80~percent of the stars formed before 8~Gyr ago. However, we also observe star formation as recent as $\sim$1~Gyr. While the stars younger than 5~Gyr only represent about 10~percent of the total mass of stars ever formed in the observed fields, they represent 20--25~percent of the most metal-rich stars. According to our SFH, we estimate that over the combined area of the \emph{SWEEPS+Stanek+Baade} fields ($\sim 120\ {\rm pc^2}$), and taking into account the $\sim$22~percent completeness fraction of bulge stars in the \emph{cleanest} sample (see Appendix~\ref{A1}), a total stellar mass of 5$\times 10^5\ M_{\sun}$ formed over 14~Gyr. The stars younger than 5~Gyr represent 6~percent of that mass, or about 3$\times 10^4\ {\rm M_{\sun}}$.
Considering only the stars that are still alive today and within reach of the current generation of spectrographs (i.e.\ $V\la$~21), we find that 10~percent of the bulge stars are younger than 5~Gyr.

One of the outputs of our SFH calculations is the AMR, from which we can also extract the present-day MDF. The AMR indicates a fast chemical enrichment up to [Fe/H]$\sim$0.4 about 7~Gyr ago corresponding to a rate of $dZ/dt \sim 0.005\ {\rm Gyr^{-1}}$, but little evolution since then. Our MDF accurately reproduces the MDF obtained by the recent spectroscopic surveys of Baade's window, with the bulk of stars having metal-content in the range [Fe/H]$\sim-$0.7 to $\sim$0.6 and a sparse tail to much lower metallicities.

\section*{Acknowledgements}

The authors are grateful to the anonymous referee for insightful comments that helped improve the presentation of the paper and strengthen our results.
EJB acknowledges support from the CNES postdoctoral fellowship program.
This research work has made use of the Python packages Numpy\footnote{http://www.numpy.org/} \citep{wal11},
Astropy\footnote{http://www.astropy.org} \citep{ast13}, and Matplotlib\footnote{http://matplotlib.org/} \citep{hun07}.


%%%%%%%%%%%%%%%%%%%%%%%%%%%%%%%%%%%%%%%%%%%%%%%%%%

\appendix

\section{Biases from the proper-motion cleaning}\label{A1}

Given the complex nature and dynamics of the Galactic bulge populations, attempting to separate bulge member from foreground disc contamination is bound to introduce selection biases in our CMDs, and potentially affect the calculated SFHs. To estimate the impact of the PM-cuts we use the Galactic bulge model M1 presented in \citet{fra17}; we refer the interested reader to this paper for a detailed description of the simulations. In short, it is composed of three populations that originated in a thin disc, a young thick disc, and an old thick disc, each with their own spatial and dynamical properties and specific distribution of ages and metallicities.

\begin{figure}
 \includegraphics[width=8.5cm]{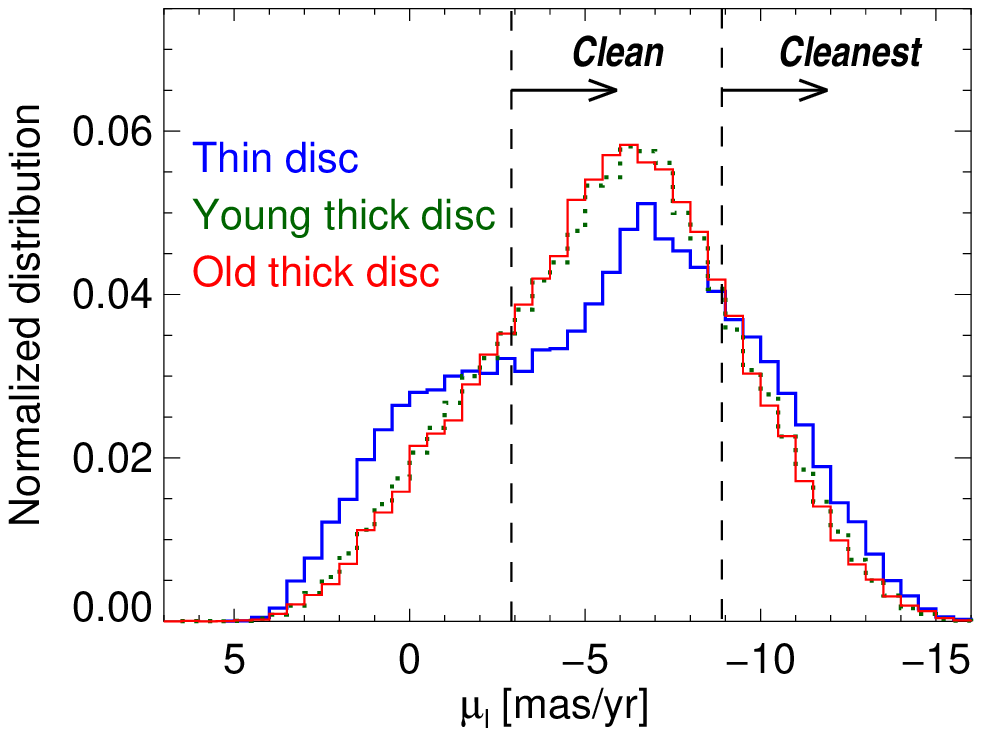}
 \caption{Longitudinal PM distribution in the bulge model of \citet{fra17}, where each component is represented by a different linestyle: thin disc in solid blue, young thick disc in dashed green, and old thick disc in solid red. The cuts corresponding to the \emph{clean} and \emph{cleanest} samples are shown at $\pm 3\ {\rm mas\ yr}^{-1}$ from the median PM of the thick disc component ($\overline{\mu_l} = -5.9\ {\rm mas\ yr}^{-1}$).}
 \label{figA1}
\end{figure}

\begin{figure}
 \includegraphics[width=8.5cm]{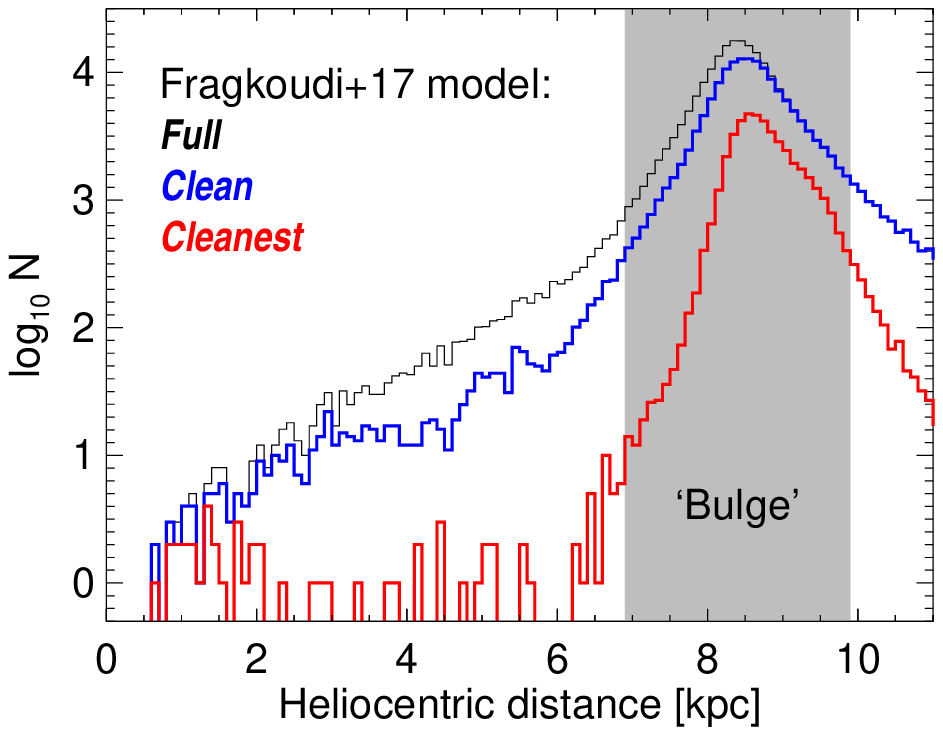}
 \caption{Density of stars along the line of sight to the bulge, as a function of heliocentric distance. The gray shaded area represents the volume within 1.5~kpc from the center of the model galaxy, which we associate with the bulge. The blue and red lines correspond to our \emph{clean} and \emph{cleanest} samples, respectively.}
 \label{figA2}
\end{figure}

While the simulation covers the whole Galaxy we selected the stars within a box including our \emph{HST} fields ($2\degr > l > -0.5\degr$, $-1.5\degr > b > -4.5\degr$), then applied the same cuts as for our \emph{clean} and \emph{cleanest} samples, i.e.\ at $\pm 3\ {\rm mas\ yr}^{-1}$ from the median PM of the thick disc component ($\overline{\mu_l} = -5.9\ {\rm mas\ yr}^{-1}$). This is illustrated in Figure~\ref{figA1}, which presents the distribution of longitudinal PM of each bulge component, and the cuts as vertical dashed lines.
We note that the PM are \emph{not} centered on zero like the observed PM shown in Figure~\ref{fig05} since the latter are relative to the median PM.
Figure~\ref{figA2} shows that these cuts are very efficient at removing the foreground disc stars: the \emph{clean} cut retains 74.6~percent of bulge stars and 39.7~percent of disc stars, while the \emph{cleanest} cut retains 21.8~percent of bulge stars but only 1.2~percent of the disc contaminants.

\begin{figure}
 \includegraphics[width=8.5cm]{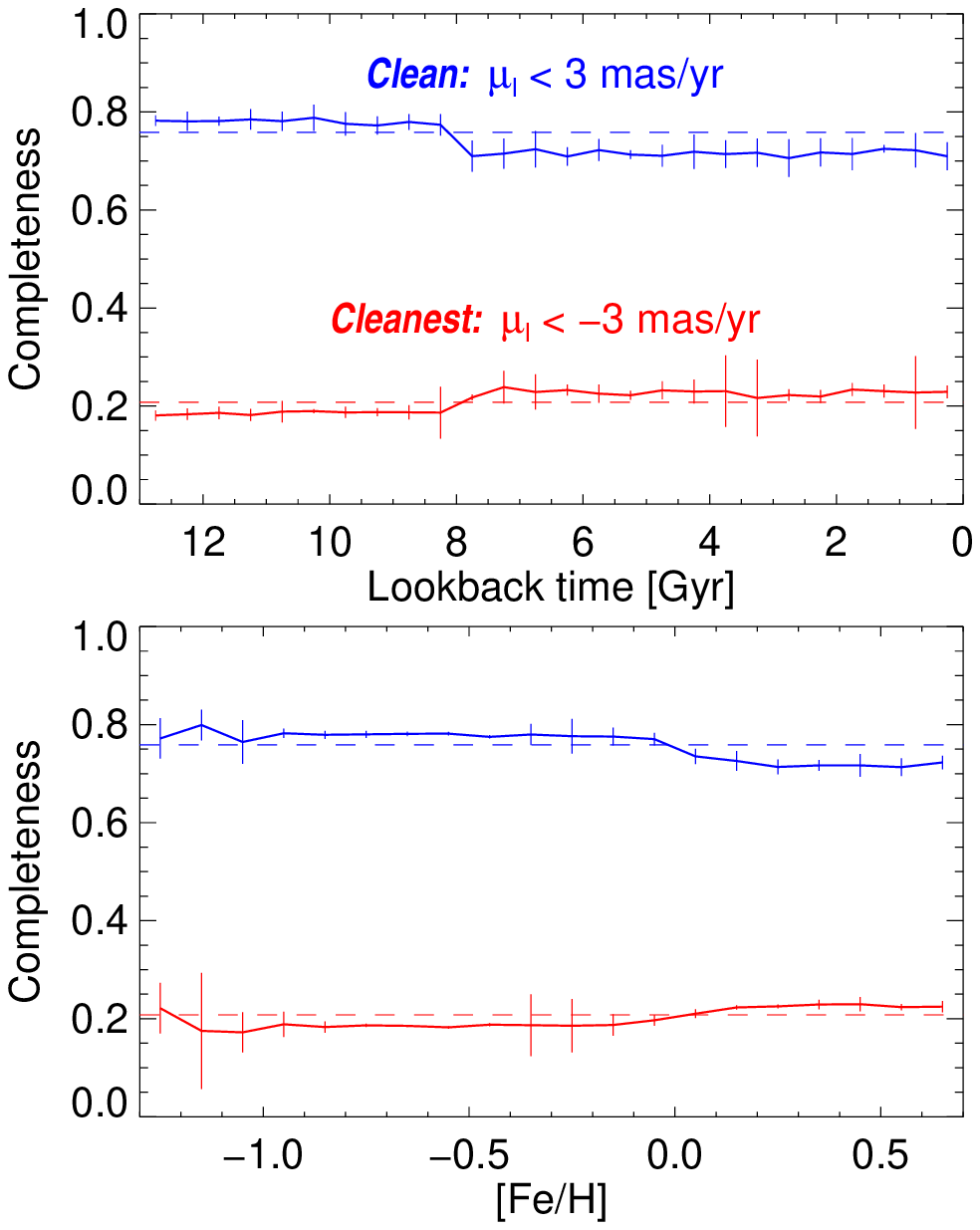}
 \caption{Variation of the completeness fraction as a function of age (top) and metallicity (bottom) as a consequence of the PM cuts. The blue and red lines correspond to our \emph{clean} and \emph{cleanest} samples. The transition from the thick to thin disc populations is visible at 8~Gyr and [Fe/H]=0.}
 \label{figA3}
\end{figure}

Figure~\ref{figA3} shows the consequence of these cuts on the completeness fraction as a function of age (top) and metallicity (bottom). On average, the cuts corresponding to the \emph{clean} and \emph{cleanest} subsamples lead to 74 and 21~percent completeness, in excellent agreement with the fractions for the real observations (74 and 23~percent). On the other hand, the curves shown in Figure~\ref{figA3} reveal small biases varying as a function of both age and metallicity.
The \emph{clean} cut in particular biases more strongly against the young and more metal-rich component corresponding to the thin disc, with a completeness of 72~percent compared to 78~percent for the thick disc component.
As expected from the similar distribution of PM at more negative $\mu_l $ shown in Figure~\ref{figA1}, the result of the \emph{cleanest} cut is much less biased. We estimate a completeness of 23 and 19~percent for the populations that originated in the thin and thick discs, respectively, or about 1-$\sigma$ from the mean completeness. We therefore conclude that the \emph{cleanest} PM-cuts have a negligible bias on our calculated SFH.

\section{CMD residuals}\label{B1}

\begin{figure*}
\includegraphics[width=5.8cm]{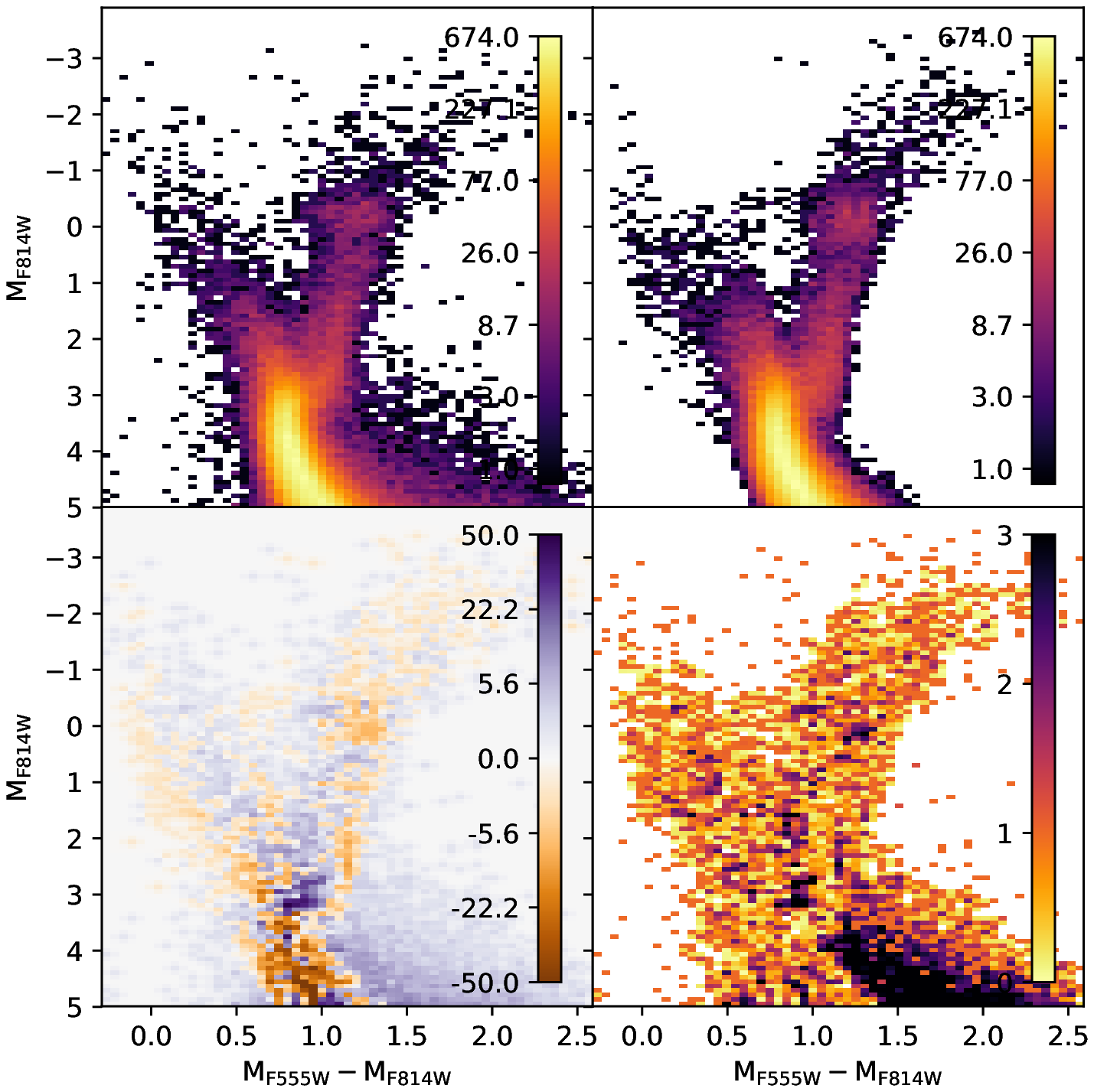}
\includegraphics[width=5.8cm]{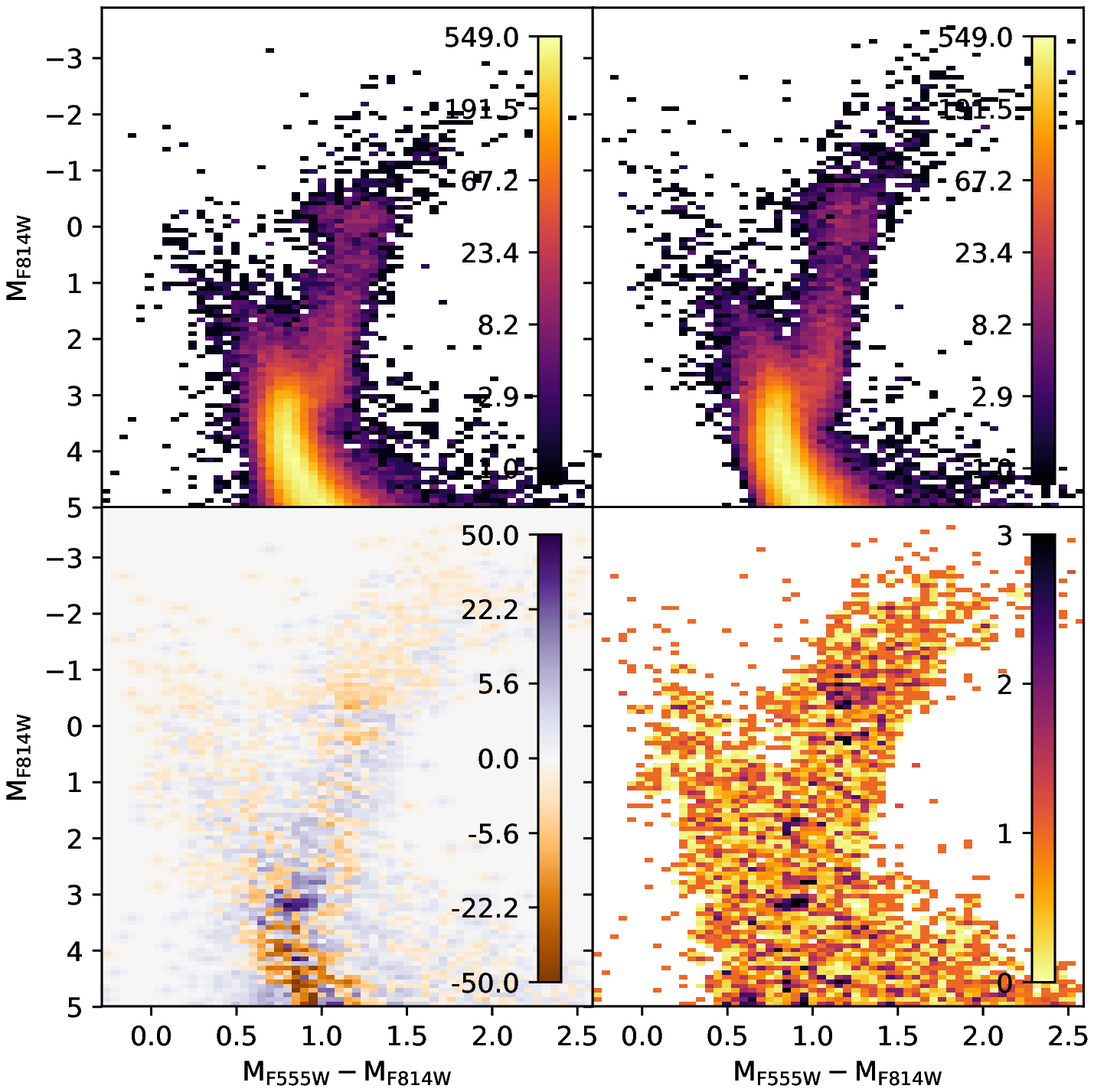}
\includegraphics[width=5.8cm]{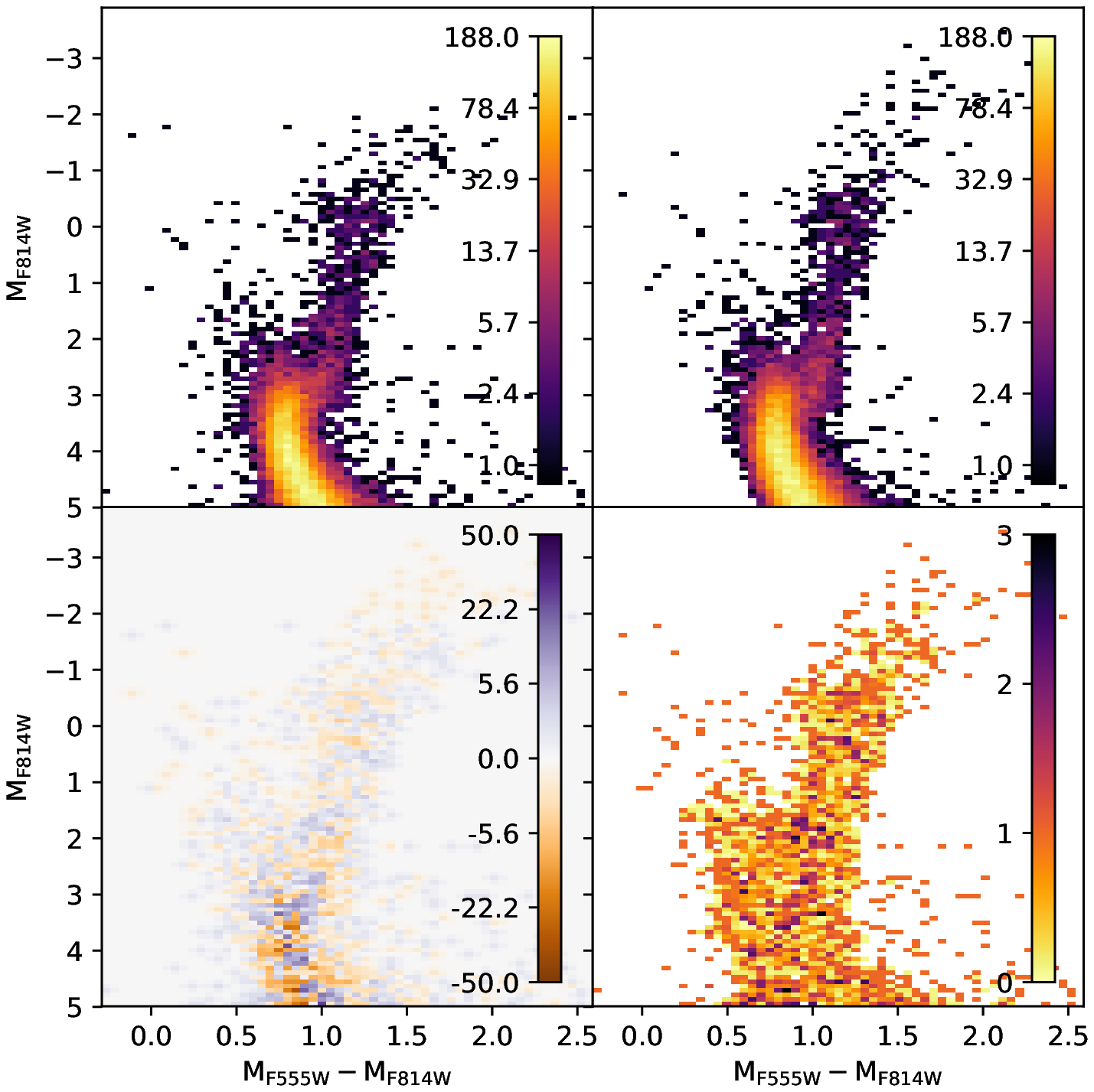}
\caption{Comparison of the observed and best-fit CMDs for the full (left), clean (middle), and cleanest (right) samples. In each case, clockwise the four panels represent: the observed CMD, the best-fit CMD, the difference in Poissonian sigmas, and the difference in number of stars per bin.}
\label{figB1}
\end{figure*}

Figure~\ref{figB1} presents the comparison between the observed and best-fit CMDs for the combined \emph{SWEEPS+Stanek+Baade} fields. For each cleaning level, the panels correspond to (clockwise): the observed CMD, the best-fit CMD, the difference in Poissonian sigmas, and the difference in number of stars per bin. The residuals for the `full' sample (left) are significant, since we did not attempt to fit the contamination with our foreground population. On the other hand, for the two other samples the residuals are very low and do not show any significant, coherent structure.

\section{Individual SFHs}\label{C1}

\begin{figure*}
\includegraphics[width=5.8cm]{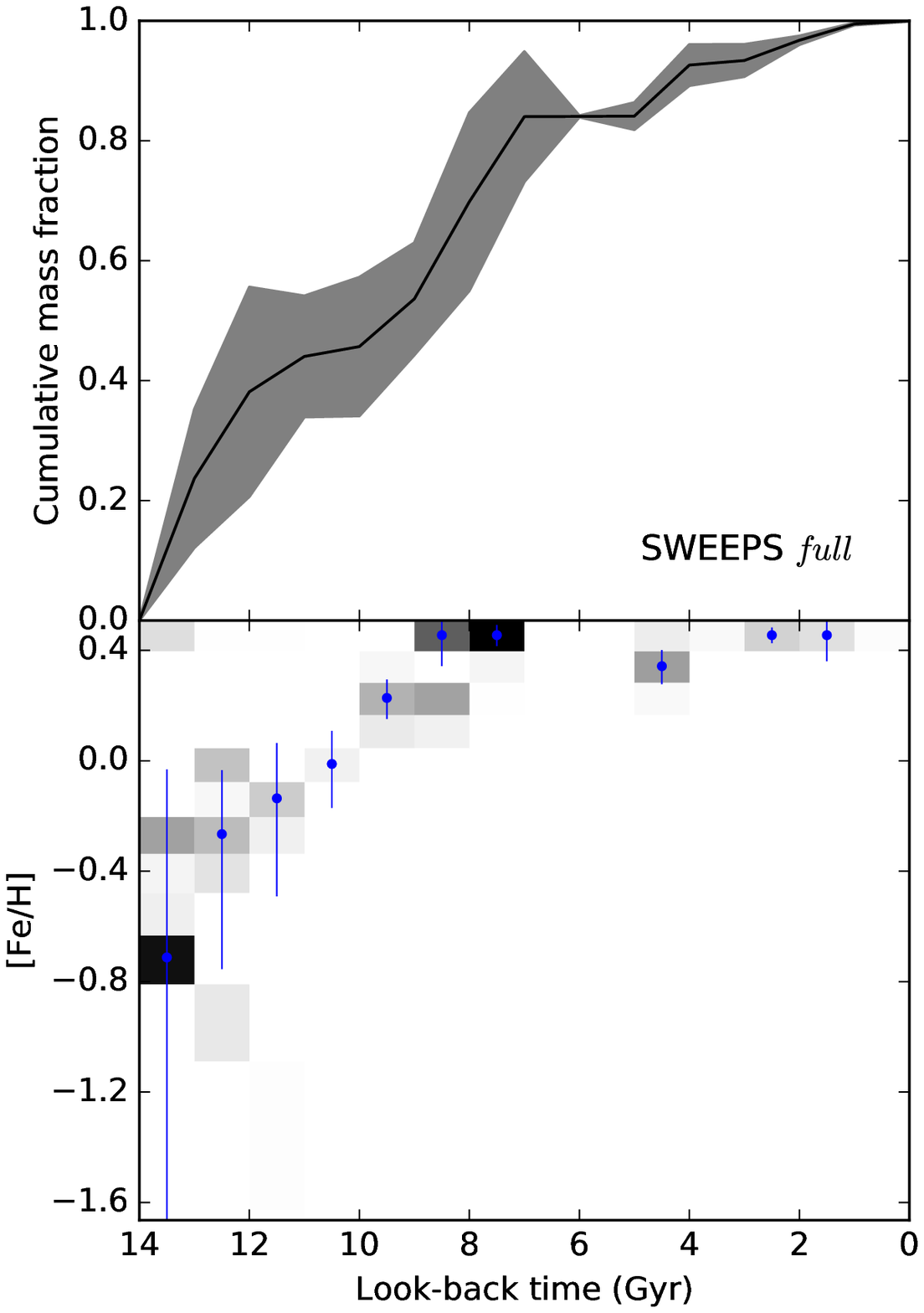}
\includegraphics[width=5.8cm]{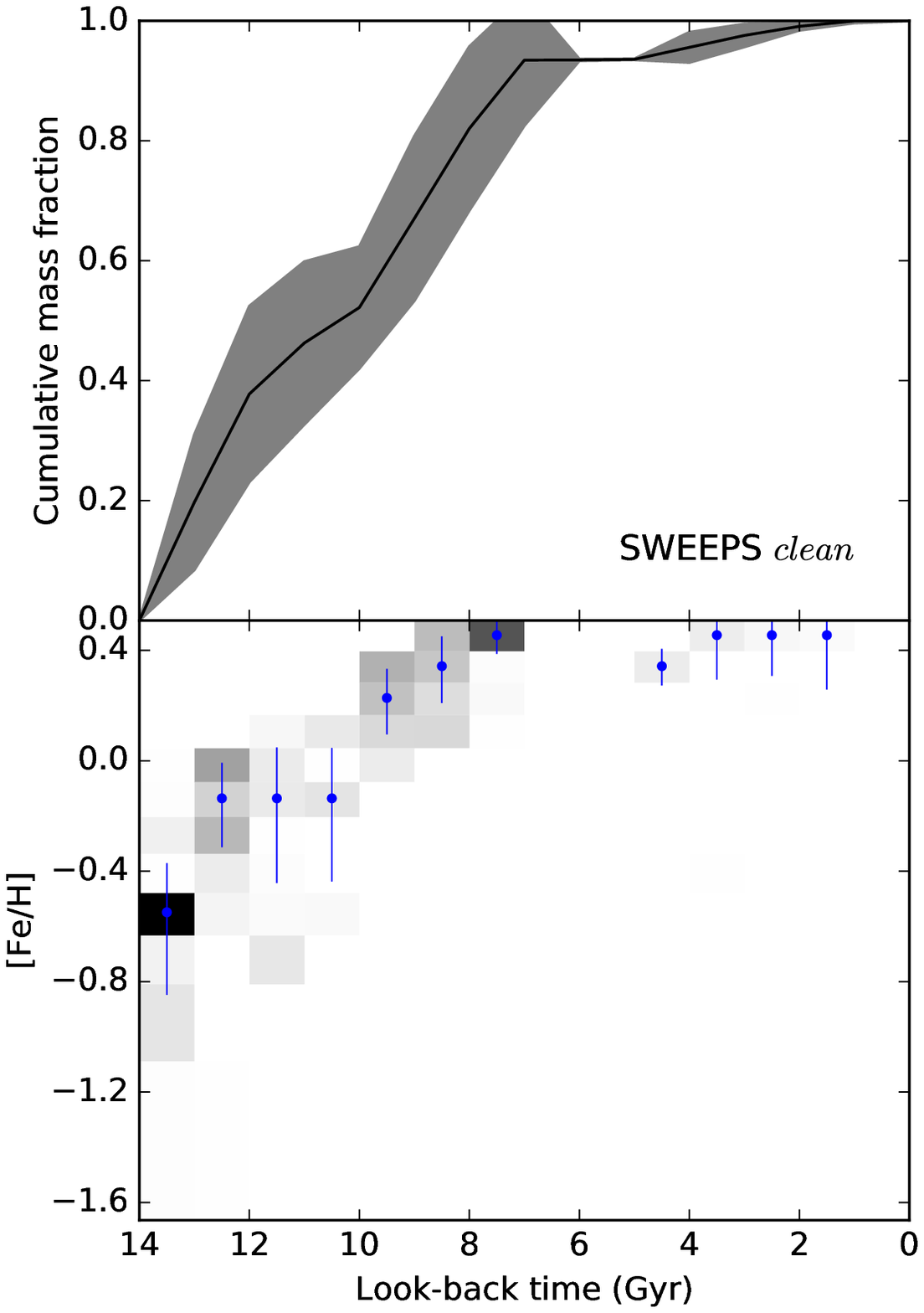}
\includegraphics[width=5.8cm]{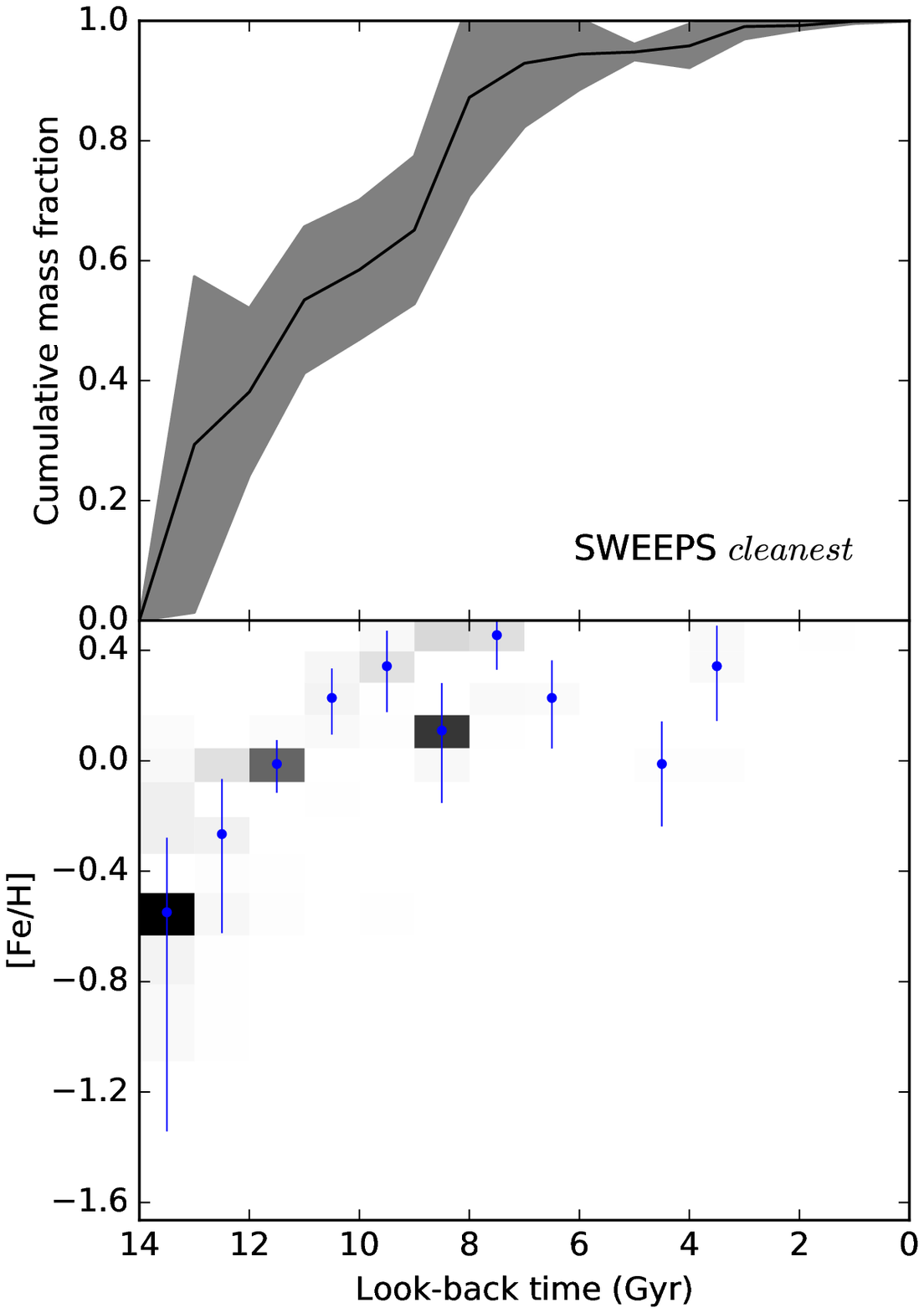}
\caption{SFH of the \emph{SWEEPS} field with the full (left), clean (middle), and cleanest (right) CMDs.}
\label{figC1}
\end{figure*}

The SFH of the three individual fields (\emph{SWEEPS, Stanek}, and \emph{Baade}) and three levels of PM-cleaning are shown in Figures~\ref{figC1}--\ref{figC3}. In all cases, we find the same trend of decreasing fraction of young stars with stricter PM-cleaning, and the same AMRs increasing from [Fe/H]$\sim-$0.5 to super-solar metallicity between 14 and 7~Gyr ago. This motivated the decision to present the analysis of the combined CMDs in the main text of the paper.

\begin{figure*}
\includegraphics[width=5.8cm]{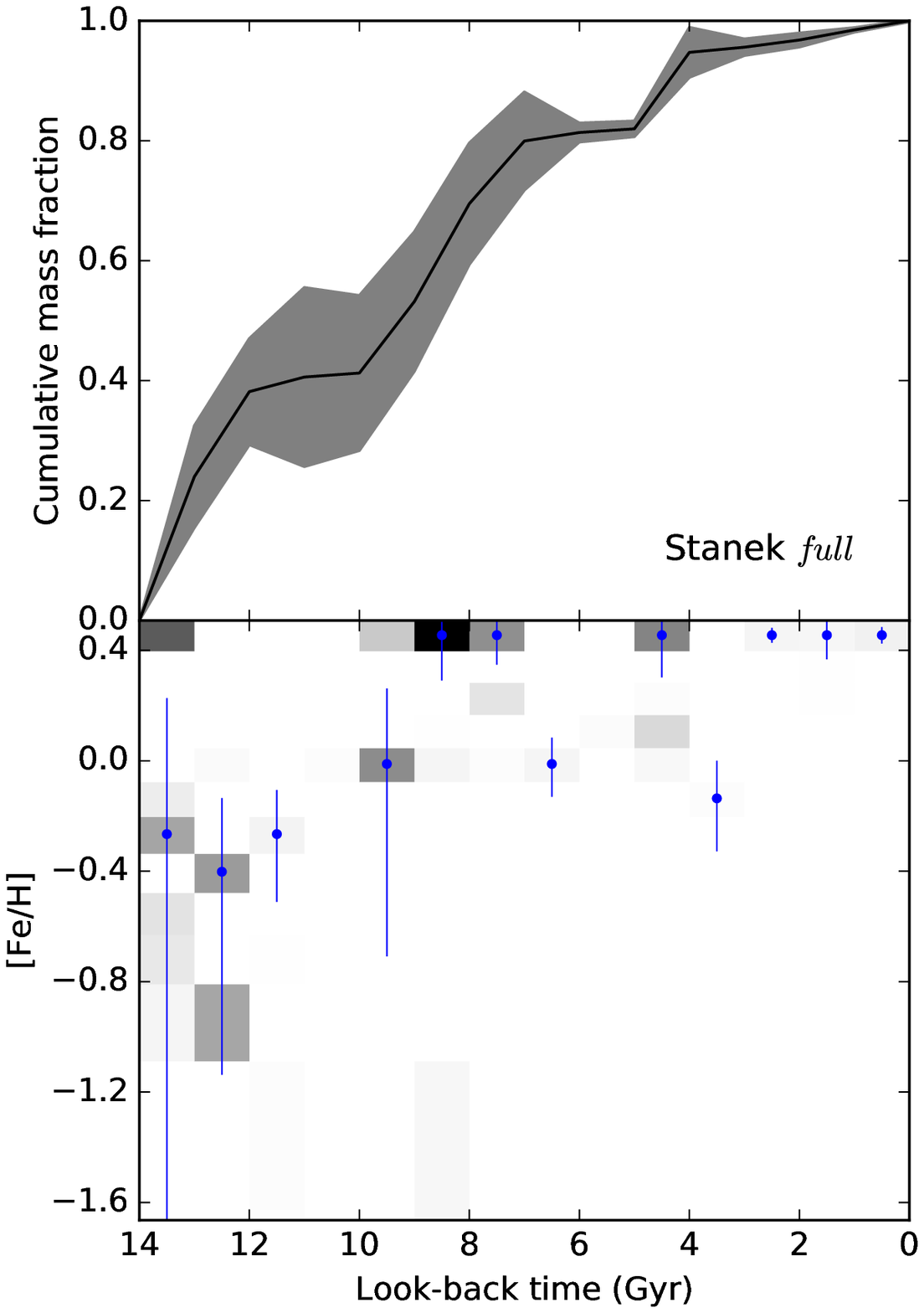}
\includegraphics[width=5.8cm]{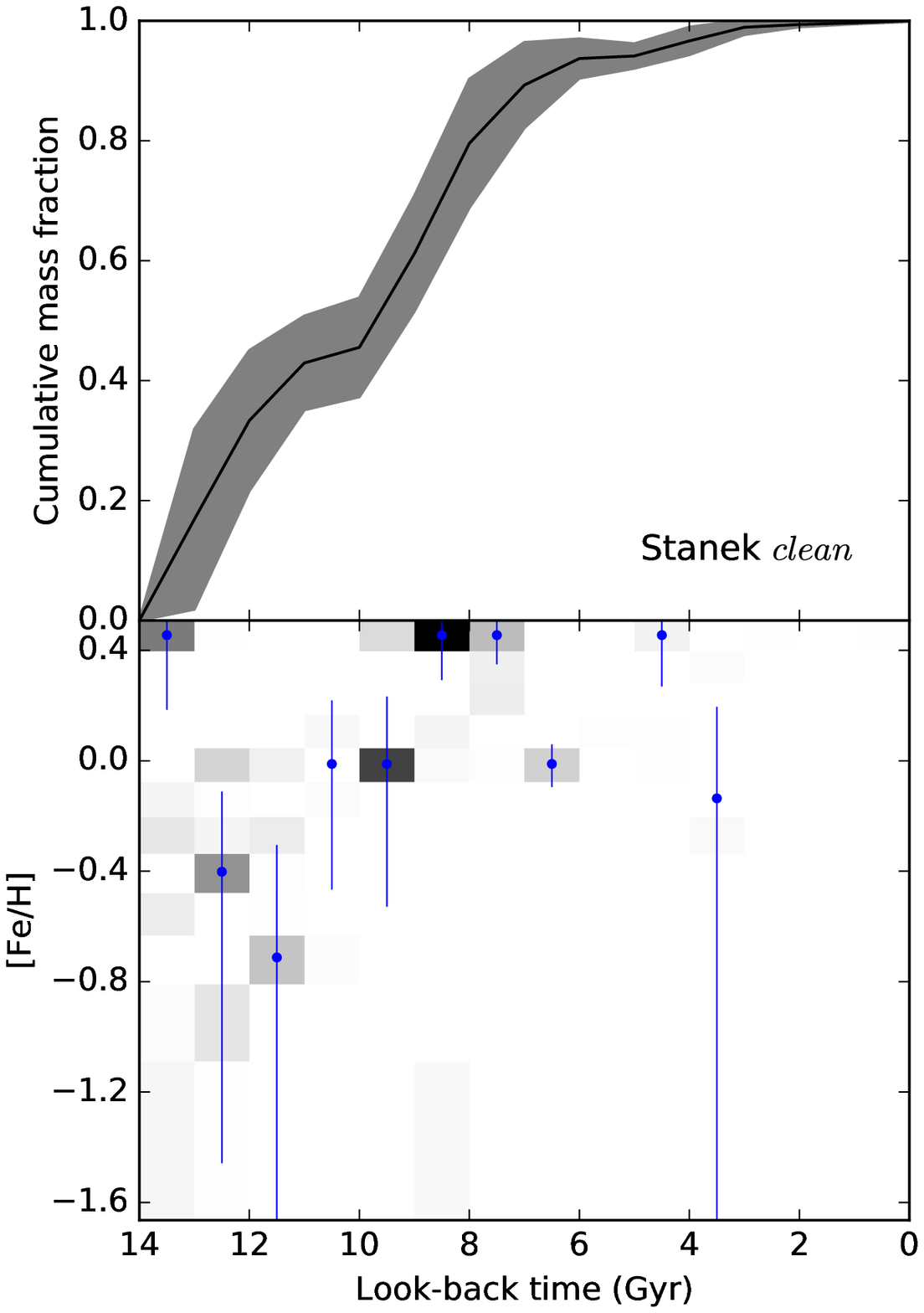}
\includegraphics[width=5.8cm]{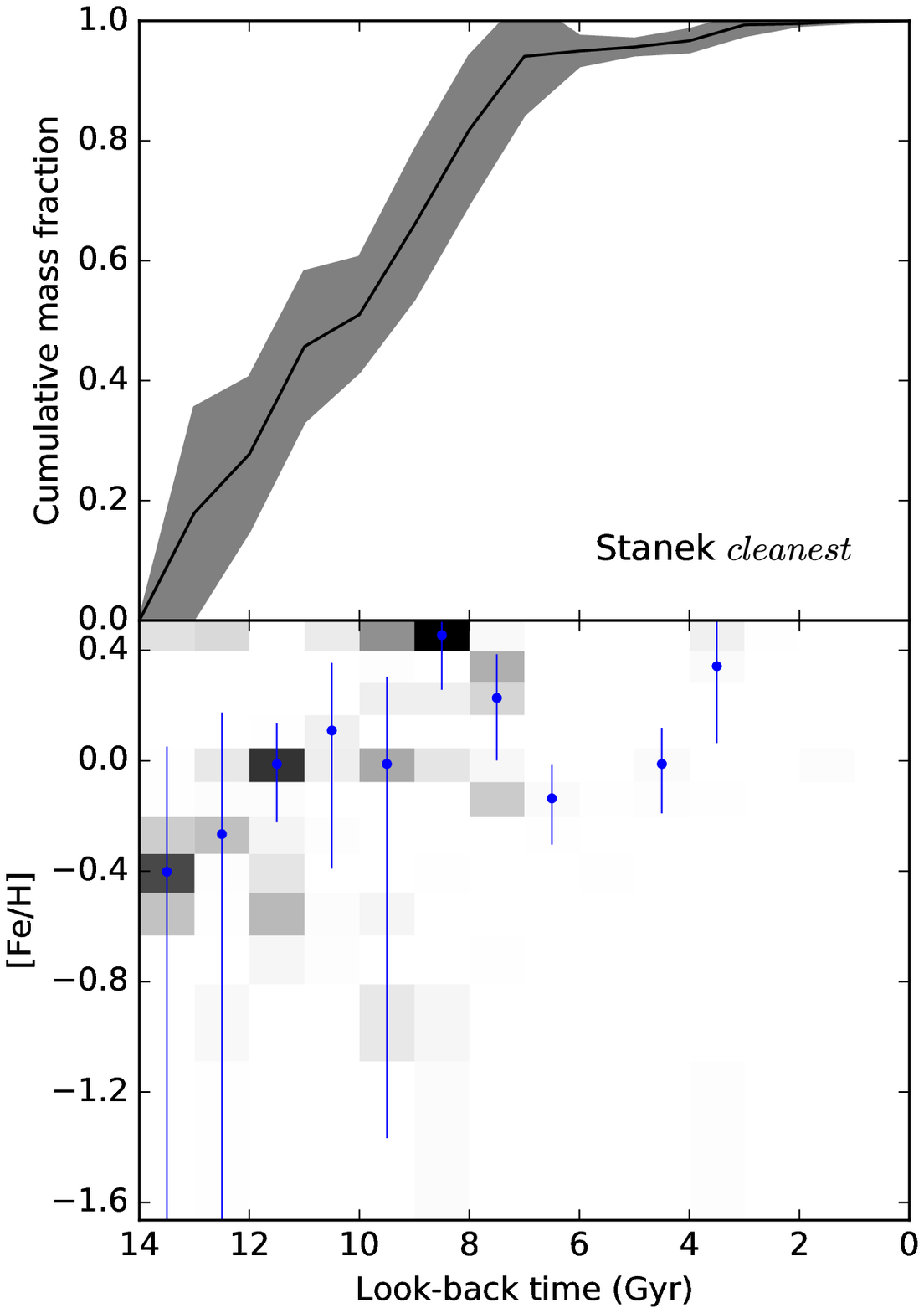}
\caption{SFH of the \emph{Stanek} field with the full (left), clean (middle), and cleanest (right) CMDs.}
\end{figure*}

\begin{figure*}
\includegraphics[width=5.8cm]{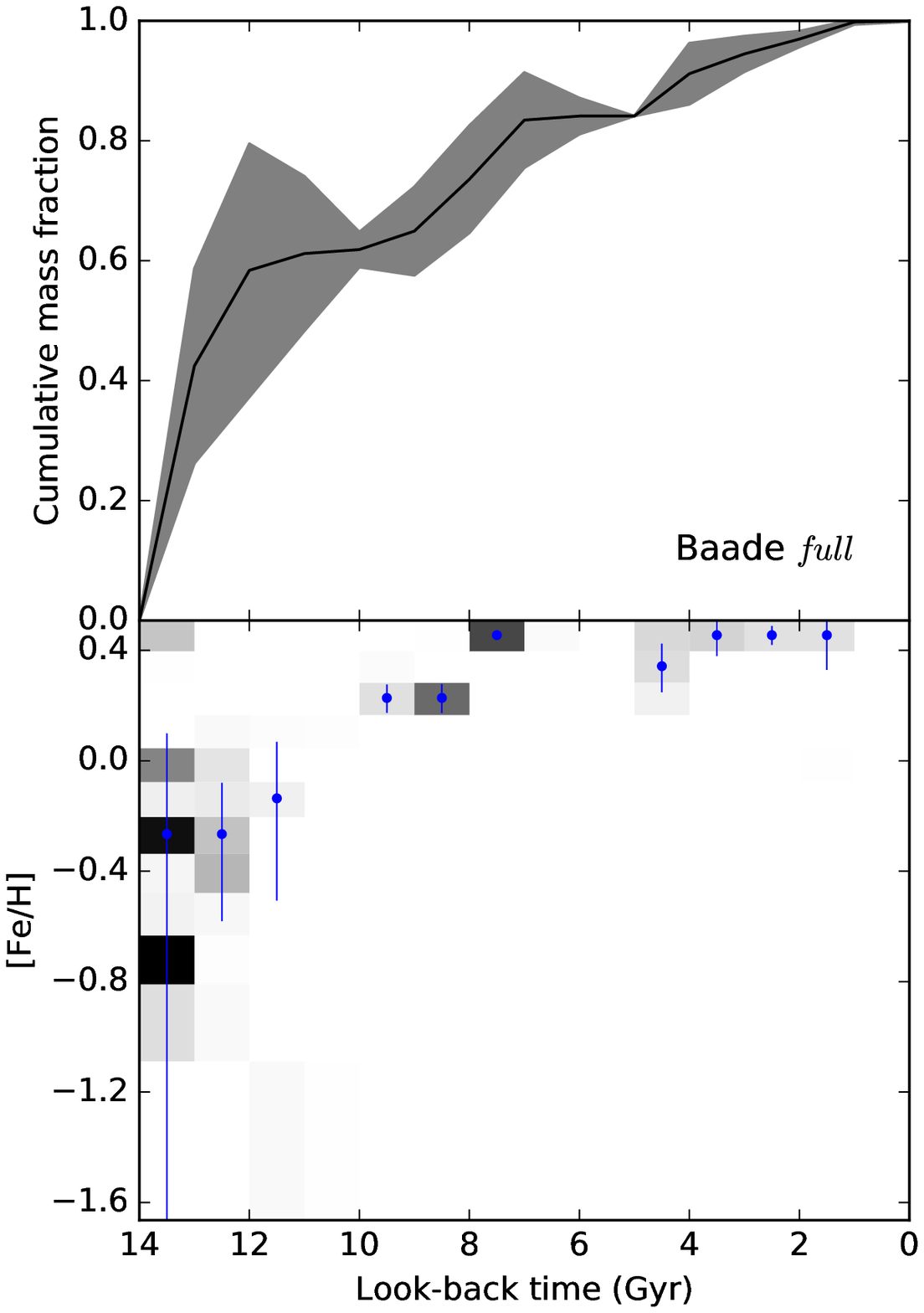}
\includegraphics[width=5.8cm]{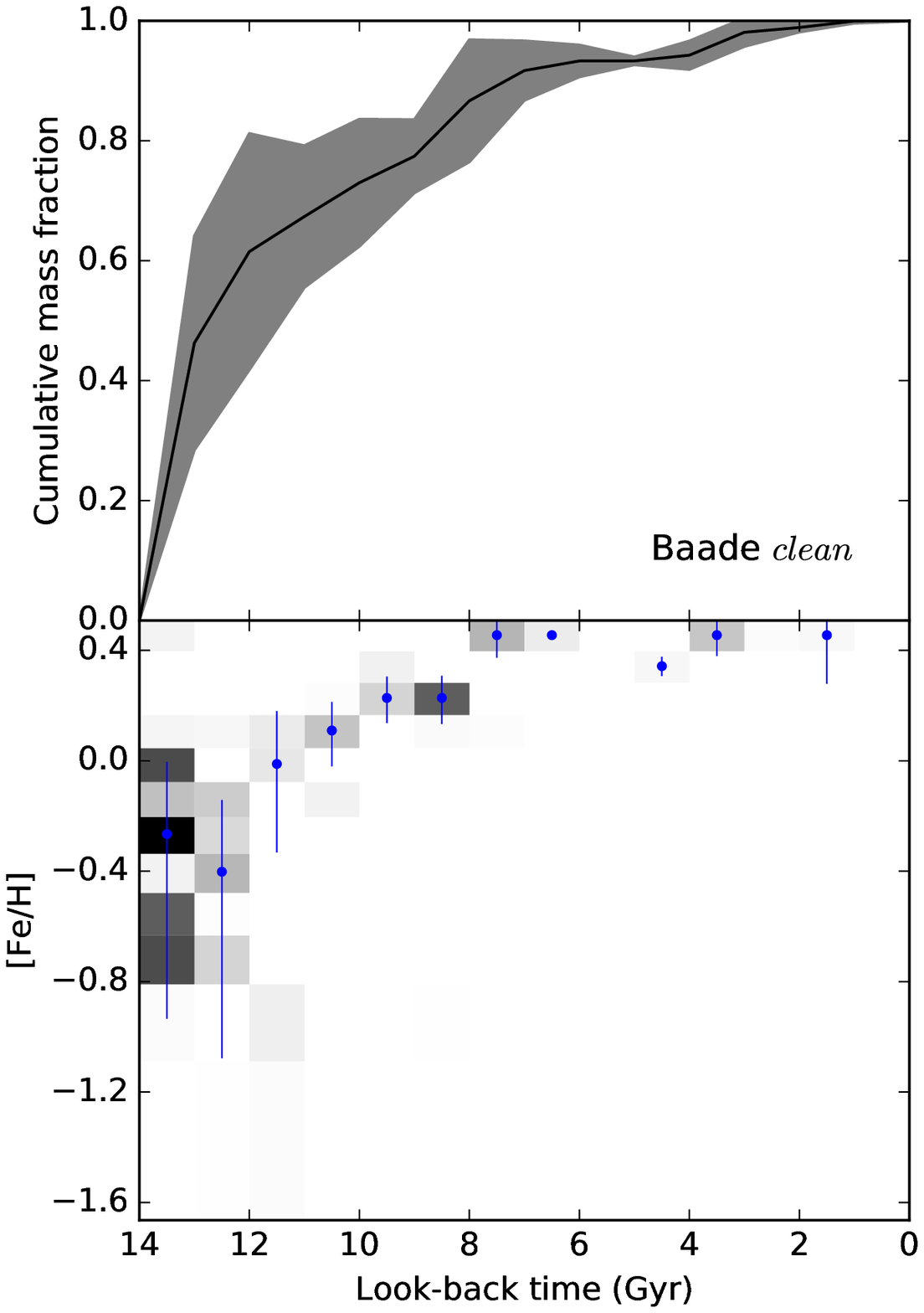}
\includegraphics[width=5.8cm]{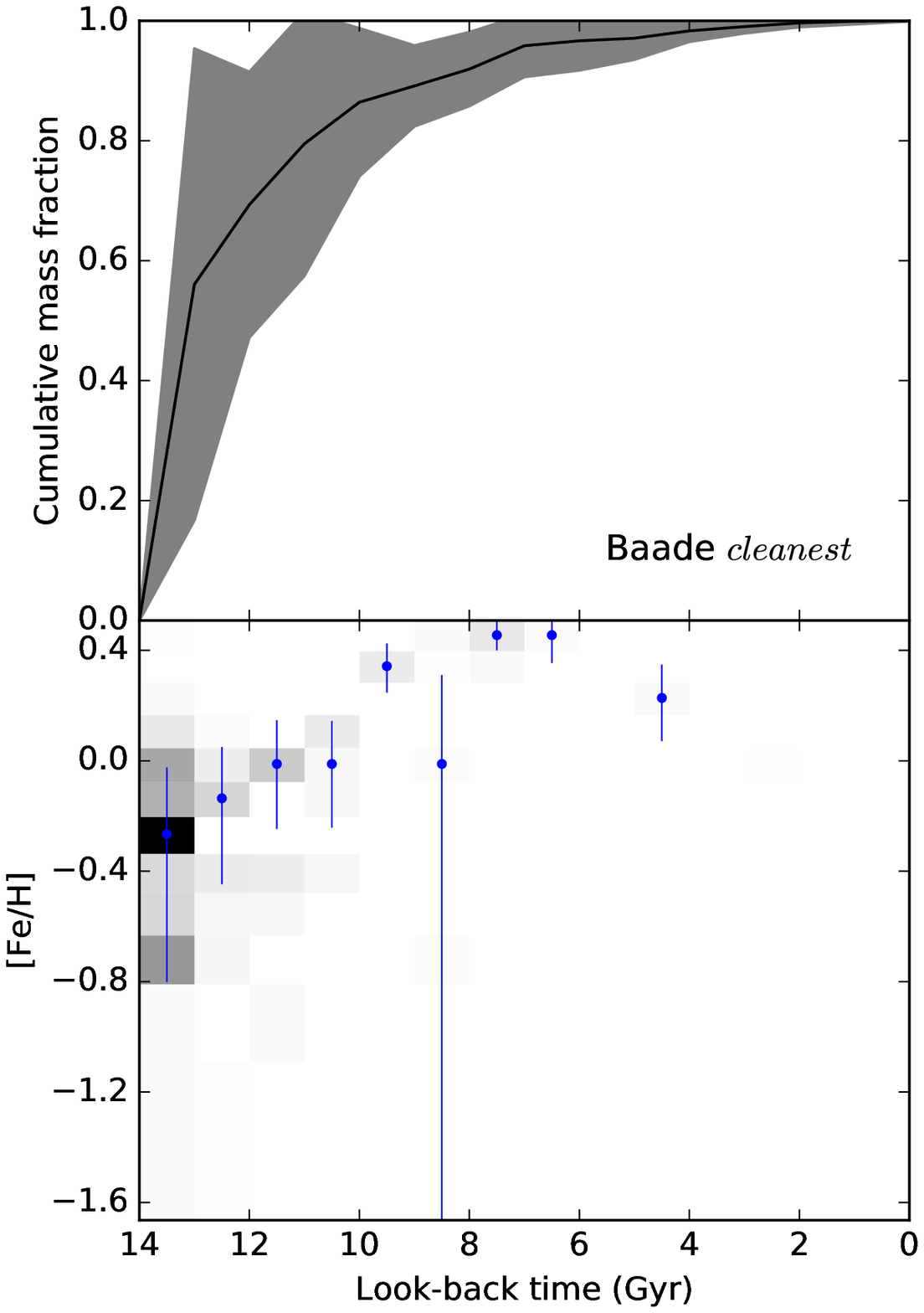}
\caption{SFH of the \emph{Baade} field with the full (left), clean (middle), and cleanest (right) CMDs.}
\label{figC3}
\end{figure*}

% Don't change these lines
% \bsp    % typesetting comment
\label{lastpage}
\end{document}